\documentclass[twocolumn, journal]{IEEEtran}

\IEEEoverridecommandlockouts

\usepackage{diagbox}
\usepackage{bm}
\usepackage{color}
\usepackage{graphicx}
\usepackage{amsmath}
\usepackage{amssymb}
\usepackage{algorithm}
\usepackage{algorithmic}
\usepackage{ulem}
\usepackage{lineno}
\usepackage{lettrine}
\usepackage{subfigure}
\usepackage{epstopdf}
\usepackage{stfloats}
\usepackage{color}
\usepackage{graphicx}
\usepackage{amsmath}
\usepackage{amssymb}
\usepackage{algorithm}
\usepackage{algorithmic}
\usepackage{amsmath}
\usepackage{multirow}
\usepackage{booktabs}
\usepackage{array}
\usepackage{amsthm}
\usepackage{stfloats}
\usepackage{caption}
\usepackage{subfigure}
\usepackage{bm}
\usepackage{booktabs}
\usepackage{setspace}
\usepackage{makecell}
\usepackage{cases}

\newcommand{\be}{\begin{equation}}
\newcommand{\ee}{\end{equation}}
\newcommand{\bea}{\begin{eqnarray}}
\newcommand{\eea}{\end{eqnarray}}
\newcommand{\ba}{\begin{array}}
\newcommand{\ea}{\end{array}}
\newcommand{\nid}{\noindent}

\renewcommand{\thesubsubsection}{\thesubsection.\arabic{subsubsection}}

\allowdisplaybreaks[4]

\title{Low-Complexity Designs of Symbol-Level Precoding for MU-MISO Systems
\thanks{Part of this paper is presented in IEEE Global Communications Conference (GLOBECOM), 2021 \cite{Xiao Globecom 2021}.}
\thanks{ Z. Xiao, R. Liu, M. Li, and Y. Liu are with the School of Information and Communication
Engineering, Dalian University of Technology, Dalian 116024, China (e-mail:
xiaozichao@mail.dlut.edu.cn; liurang@mail.dlut.edu.cn; mli@dlut.edu.cn; yangliu\_613@dlut.edu.cn).}
\thanks{ Q. Liu is with the School of Computer Science and Technology, Dalian University
of Technology, Dalian 116024, China (e-mail: qianliu@dlut.edu.cn).}}

\author{Zichao Xiao,
        Rang Liu,~\IEEEmembership{Graduate Student Member,~IEEE,}
        Ming Li,~\IEEEmembership{Senior Member,~IEEE,}\\
        Yang Liu,~\IEEEmembership{Member,~IEEE,}
        and Qian Liu,~\IEEEmembership{Member,~IEEE}
}

\begin{document}
\maketitle
\thispagestyle{empty}
\begin{abstract}
Symbol-level precoding (SLP), which converts the harmful multi-user interference (MUI) into beneficial signals, can significantly improve symbol-error-rate (SER) performance in multi-user communication systems.
While enjoying symbolic gain, however, the complicated non-linear symbol-by-symbol precoder design suffers high computational complexity exponential with the number of users, which is unaffordable in realistic systems.
In this paper, we propose a novel low-complexity grouped SLP (G-SLP) approach and develop efficient design algorithms for typical max-min fairness and power minimization problems.
In particular, after dividing all users into several groups, the precoders for each group are separately designed on a symbol-by-symbol basis by only utilizing the symbol information of the users in that group, in which the intra-group MUI is exploited using the concept of constructive interference (CI) and the inter-group MUI is also effectively suppressed.
In order to further reduce the computational complexity, we utilize the Lagrangian dual, Karush-Kuhn-Tucker (KKT) conditions and the majorization-minimization (MM) method to transform the resulting problems into more tractable forms, and develop efficient algorithms for obtaining closed-form solutions to them.
Extensive simulation results illustrate that the proposed G-SLP strategy and design algorithms dramatically reduce the computational complexity without causing significant performance loss compared with the traditional SLP schemes.
\end{abstract}


\section{Introduction}

Over the last decade, multiple-input multiple-output (MIMO) technique with multi-antenna arrays has become a key enabling technique for wireless communication systems \cite{MIMO1}, \cite{MIMO2}, such as 5G and beyond networks, which significantly improves the system performance in terms of spectrum efficiency, energy efficiency, coverage, user capacity, etc.
Meanwhile, precoding has been playing an indispensable role in existing MIMO wireless communication systems.
Particularly, various linear precoding techniques have been developed and employed in multi-user systems to suppress or even eliminate multi-user interference (MUI) at the receiver side.
These designs usually utilize the second-order statistics of the signals as the performance metrics, e.g., maximum signal-to-interference-plus-noise ratio (SINR), minimum mean square error (MMSE), to find a linear mapping between symbols and precoders, by which a block of symbols can be easily precoded by low-complexity block-level precoding (BLP) \cite{precoding1}, \cite{precoding2}.

Unlike the traditional linear BLP technique which aims at suppressing or eliminating MUI, recently emerged symbol-level precoding (SLP) technique treats MUI as a source of useful signals that can enhance the system performance  \cite{Tutorial1}, \cite{Tutorial2}.
Specifically, by exploiting the knowledge of channel state information (CSI) and transmitted symbol information, the precoder at each time-slot is elaborately designed to convert harmful MUI into useful signals via the concept of constructive interference (CI).
This non-linear symbol-to-precoder mapping exploits both the spatial and symbol-level degrees of freedom (DoFs) for optimizations and consequently achieves significantly better performance compared with BLP schemes.
Therefore, in recent years there is a growing interest in exploring the potential advantages of SLP technique for various wireless application scenarios.

The initial works \cite{SLP performance2}, \cite{SLP performance} verified the superiority of this symbol-to-precoder scheme in significantly reducing the symbol-error-rate (SER) of multi-user multi-input single-output (MU-MISO) systems by exploiting MUI.
With the well-established model of SLP for MU-MISO systems, various practical hardware limitations have been taken into account, such as the constant envelope architecture with low peak-to-average-power ratio (PAPR) constraint \cite{massive MIMO CE}, \cite{massive MIMO1}, the low-resolution digital-to-analog converter (DAC) \cite{massive MIMO3}, the hybrid analog-digital architecture \cite{massive MIMO Hybrid1}, \cite{massive MIMO Hybrid2}, and the single radio frequency (RF), RF-domain architecture \cite{massive MIMO4}, etc.
In addition to enjoying the superiority of SLP in reducing SER or transmit power for MU-MISO systems, researchers have also devoted themselves to combining SLP with other techniques, such as simultaneous wireless information and power transfer (SWIPT) \cite{SLP SWIPT1}, \cite{SLP SWIPT3}, cognitive radio (CR) \cite{SLP CR1}, \cite{SLP CR2}, faster-than-Nyquist (FTN) signaling \cite{FTN}, multi-cell scenario \cite{CI-based multi-cell}, physical layer security (PLS) \cite{Rang Secure2}, \cite{Xu TIFS 2020}, intelligent reflecting surfaces (IRS) \cite{IRS SLP1}-\cite{IRS SLP3}, integrated sensing and communication (ISAC) \cite{RAC1}, \cite{RAC2}, to take the advantages of the temporal-domain flexibility and the CI.

While benefiting the significant advancements of SLP in exploiting MUI, however, the non-linear symbol-by-symbol precoder design causes dramatically higher computational complexity compared with its BLP counterparts.
Specifically, since the SLP technique optimizes the transmit precoder for each specific transmitted symbol vector, the total number of the precoders to be designed during a channel coherent time equals to the number of different possible transmitted symbol vectors, which is exponential with the number of users.
The huge number of required optimizations prevents the applications of SLP technique in the systems with large numbers of users.
Ironically, more users may potentially generate more MUI to be exploited and consequently using SLP in dense-user systems can offer much more performance improvement over conventional BLP.
Therefore, a breakthrough in reducing the complexity of SLP designs is vitally important for benefiting from SLP in practical dense-user systems.

In recent years, low-complexity SLP designs have drawn increasing research attentions.
One popular approach is to simplify the original optimization problem and derive closed-form solutions \cite{A Li CF}-\cite{PM ICF}.
Particularly, the authors in \cite{A Li CF} simplified the max-min fairness problem for phase-shift-keying (PSK) modulation by using the Lagrangian dual and Karush-Kuhn-Tucker (KKT) conditions, and then proposed an iterative algorithm with conditionally optimal closed-form solutions.
The extension to quadrature amplitude modulation (QAM) signalling was further investigated in \cite{A Li CF QAM}.
For the typical power minimization problem, the authors in \cite{Low complexity PM} analyzed the structure of the optimal solution based on the KKT conditions, and then derived a suboptimal closed-form solution.
This approach was further improved in \cite{PM ICF} where the conditions for nearly perfect recovery of the optimal solution support was utilized.
In addition to the above optimization-based methods, machine learning based methods have also been investigated to tackle this puzzle \cite{Deep learning1}-\cite{Deep learning3}.
An auto-encoder based deep learning network was proposed in \cite{Deep learning1} for SLP and symbol detection designs with imperfect CSI.
Unsupervised learning based precoding networks for solving the max-min fairness and power minimization problems were respectively developed in \cite{Deep learning2} and \cite{Deep learning3}.

While above mentioned optimization-based and learning-based low-complexity SLP designs greatly reduce the complexity of designing each precoder, the total number of precoders to be designed still remains enormous, especially for the systems with a large number of users.
Therefore, the major obstacle of reducing the overall complexity and implementing SLP is how to reduce the number of precoders required to be optimized.
A positive attempt in this aspect was firstly made in \cite{SLP rotation}, where the authors exploited the symmetry characteristic of symbols and successfully reduced the number of precoders required to be designed to a quarter of the original one for QPSK signals.
However, such complexity reduction is a drop in the ocean and could not fundamentally tackle the issue.
In addition, a CI-based BLP approach was proposed in the very recent work \cite{CI-BLP}, in which a constant precoding matrix is applied for a block of symbol slots within a channel coherence interval.
While this approach enjoys lower complexity for a larger block length, the resulting performance is increasingly deteriorating, which will result in an unaffordable performance loss for practical systems with a moderately large block length.
Therefore, it requires further explorations of low-complexity SLP designs.

Motivated by these findings, in this paper we propose a novel grouped SLP (G-SLP) strategy to significantly reduce the number of required precoder designs, and then develop efficient algorithms to solve the max-min fairness and power minimization problems.
In particular, we consider an MU-MISO system where a base station (BS) utilizes the proposed G-SLP strategy to realize low-complexity precoder designs and enhance the system performance by exploiting the intra-group MUI and suppressing the inter-group MUI.
Our major contributions are summarized as follows:
\begin{itemize}
\item  We propose a novel low-complexity G-SLP strategy. Specifically, we first divide all users into several groups based on the channel characteristics and decompose the precoder designs for all users into the designs for the users in each group. Then, for the users in a certain group, the symbol-level precoders are optimized to exploit the intra-group MUI by only utilizing the symbol information of the users in that group, as well as suppress the interference caused to the users in other groups.
    It should be emphasized that although the system model is similar to the CI-based multi-cell case in \cite{CI-based multi-cell}, this work focuses on the low-complexity algorithm designs associated with the grouping strategy, the power allocation, and the inter-group interference suppression problems, which are not considered in the conventional CI-based multi-cell systems.

\item An efficient G-SLP design algorithm is developed to solve the max-min fairness problem for an MU-MISO system. We first relax the inter-group MUI suppression constraint to a more favorable linear form and utilize the Lagrangian and KKT conditions to convert the resulting problem into a much simpler one with only linear constraints.
    Then, the majorization-minimization (MM) method is employed to obtain a more tractable surrogate objective function.
    Finally, the resulting problem is split into two sub-problems, and each of them is efficiently solved in closed-form.

\item Then, an efficient G-SLP design algorithm is developed to tackle the power minimization problem.
    We first follow a similar procedure that utilizes Lagrangian, KKT conditions, and the MM method to transform the optimization problem into a more tractable one, and then conduct closed-form solutions to the resulting problems.

\item Finally, extensive simulation results are provided to verify the advantages of the proposed G-SLP strategy and the effectiveness of the developed algorithms.
    Compared with existing SLP based designs, the proposed G-SLP based algorithms dramatically reduce the computational complexity by orders of magnitude at the price of acceptable performance loss in terms of SER and transmit power.
    Meanwhile, a scalable trade-off between the achieved performance and the required complexity can be found.
\end{itemize}

\textit{Notation}: Lower-case, boldface lower-case, and upper-case letters indicate scalars, column vectors, and matrices, respectively.
$(\cdot)^T$ and $(\cdot)^H$  denote  the transpose and transpose-conjugate operations, respectively.
$\mathbb{C}$ denotes the set of complex numbers.
$\Re\{ \cdot \}$ and $\Im\{ \cdot \}$  extract the real and imaginary part of a complex number, respectively.
$\mathbf{I}_{N}$ indicates an $N \times N$ identity matrix and $\mathbf{1}$ represents the vector with all ones.
$\mathbf{0}$ represents the vector with all zeros.
$| a |$, $\| \mathbf{a} \|_2$, and $\| \mathbf{A} \|_F$ are the magnitude of a scaler $a$, 2-norm of a vector $\mathbf{a}$, and the Frobenius norm of a matrix $\mathbf{A}$, respectively.
$\angle a$ is the angle of complex-valued $a$.
$\text{diag} \{\mathbf{a}\}$ indicates the diagonal matrix whose diagonals are the elements of $\mathbf{a}$.
$\odot$ denote the Hadamard product operation.
$\mathbf{A} \succeq 0$ indicates that the matrix $\mathbf{A}$ is positive semi-definite.
$\mathbf{A}(i,j)$ denotes the element of the $i$-th row and the $j$-th column of matrix $\mathbf{A}$.
$\mathbf{A}^{\frac{1}{2}}$ denotes the matrix square root operation, which returns the principal square root of the matrix $\mathbf{A}$.

\section{System Model and Review of SLP}\label{system model}

\subsection{System Model}

We consider a downlink MU-MISO system, where a BS equipped with $N_{\text{t}}$ transmit antennas simultaneously serves $K$ single-antenna users over the flat-fading channel.
The transmitted symbol vector at the $n$-th time-slot is denoted as $\mathbf{s}[n] \triangleq \left[s_{1}[n],\ldots,s_{K}[n]\right]^T$, where each symbol is independently selected from an $\Omega$-PSK constellation ($\Omega=2,4,\ldots$).
For transferring $\mathbf{s}[n]$, the corresponding symbol-level precoder $\mathbf{x}[n] \in \mathbb{C}^{N_{\text{t}}}$ is elaborately designed for exploiting MUI and transmitted from the BS.
Then, the received signal of the $k$-th user at the $n$-th time-slot can be written as
\begin{equation}
\label{system model 1}
y_{k}[n]=\mathbf{h}_k^H\mathbf{x}[n]+n_k[n],
\end{equation}
where $\mathbf{h}_{k} \in \mathbb{C}^{N_{\text{t}}}$ represents the channel vector between the BS and the $k$-th user, and $n_k[n]\backsim\mathcal{CN}(0,\sigma^2)$ is the additive white Gaussian noise (AWGN) of the $k$-th user.
In order to focus on the low-complexity SLP designs, in this paper we assume perfect CSI is known at the BS.

\subsection{Traditional Symbol-Level Precoding Design} \label{sec: traditional SLP}

\begin{figure}[t]
\centering
\includegraphics[width = 2.6 in]{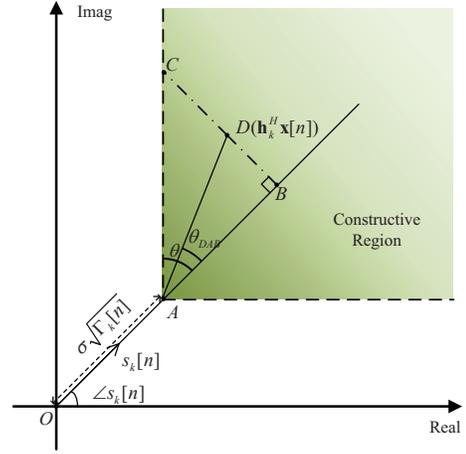}
\caption{Constructive region for a QPSK symbol.}\label{fig1} \vspace{-4 ex}
\end{figure}

With the knowledge of transmitted symbol vector $\mathbf{s}[n]$ and CSI $\mathbf{h}_k$, $k=1,\ldots,K$, MUI can be converted into beneficial signals at each user using the SLP technique.
To better understand the idea behind it, without loss of generality, we take the QPSK modulated system as an example and assume $s_{k}[n]=(\frac{1}{\sqrt{2}},j \frac{1}{\sqrt{2}})$ as the symbol of interest of the $k$-th user.
The received signal of the $k$-th user can be illustrated in a complex plane as shown in Fig. \ref{fig1}, where point $D$ represents the noise-free signal $\mathbf{h}_k^H\mathbf{x}[n]$.
At the receiver side, the decision boundaries for this symbol of interest are the positive halves of the $x$ and $y$ axes.
If the received noise-corrupted signal $y_{k}[n]$ is located in the first quadrant, the transmitted symbol can be correctly detected.
Thus, $\mathbf{x}[n]$ should be designed to let point $D$ away from the decision boundaries in order to enhance the anti-noise capability.
Since signal-to-noise ratio (SNR) is a typical metric to qualify this capability, we let $\Gamma_k[n]$ denote the SNR requirement for the $k$-th user.
If there is no MUI, for example in a single-user system, we should design the precoder to let the noise-free signal at point $A$ to guarantee that the SNR requirement is satisfied as $\frac{|\overrightarrow{OA}|^2}{\sigma^2}=\frac{|\mathbf{h}_k^H\mathbf{x}[n]|^2}{\sigma^2}=\Gamma_k[n]$. When MUI exists in multi-user systems, SLP technique aims to design precoder $\mathbf{x}[n]$ to ensure that point $D$ (i.e., $\mathbf{h}_k^H\mathbf{x}[n]$) lies in the corresponding constructive (green) region, where the MUI is utilized as beneficial components that can push the received signals away from the decision boundaries for achieving better SER performance.

To better express this design goal, we project point $D$ on the direction of $\overrightarrow{OA}$ at point $B$, define point $C$ as the intersection of the extension of $\overrightarrow{BD}$ and the nearest boundary of the constructive region, and connect points $A$ and $D$.
We denote the half of the angular range of the decision regions as $\theta=\pi/\Omega$ (i.e., $\angle CAB$ in Fig. 1), and the angle between $\overrightarrow{AD}$ and $\overrightarrow{AB}$ as $\theta_{DAB}$.
Then, we can draw the following conclusion that point $D$ in the constructive region needs satisfying $\theta_{DAB}\leq\theta$, i.e.,
\begin{equation}
\label{geometry relation}
|\overrightarrow{BC}|-|\overrightarrow{BD}| \geq 0,
\end{equation}
which is mathematically equivalent to the inequality as follows
\begin{equation}\begin{aligned}
\label{mathmatical relation}
&\big[\Re\{\mathbf{h}_k^H\mathbf{x}[n] e^{-j\angle{s_{k}[n]}}\}- \sigma\sqrt{\Gamma_k[n]}\big] \tan \theta\\
&\hspace{0.6 cm} -\big|\Im\{\mathbf{h}_k^H\mathbf{x}[n] e^{-j\angle{s_{k}[n]}}\}\big| \geq 0, \quad\forall k,~\forall n.
\end{aligned}\end{equation}
The detailed derivations are omitted here due to space limitations and readers can refer to \cite{SLP performance} and \cite{IRS SLP1} if necessary.

As mentioned above, in order to exploit MUI, we need to design the precoder $\mathbf{x}[n]$ according to the specific symbol vector $\mathbf{s}[n]$.
For the considered $\Omega$-PSK modulated system with $K$ users, there are $N_{\text{SLP}}=\Omega^K$ different possible symbol vectors to be transmitted for each coherent channel duration.
Correspondingly, the number of precoders required to be designed is $N_{\text{SLP}}$.
For conciseness, we let $\mathbf{s}_m \triangleq [s_{m,1},\ldots,s_{m,K}]^T$ denote the $m$-th kind of symbol vector, $m=1,\ldots,N_{\text{SLP}}$, and let $\mathbf{x}_m \in \mathbb{C}^{N_{\text{t}}}$ denote its corresponding precoder.
Thus, the max-min fairness optimization problem for designing $\mathbf{x}_m$ can be formulated as
\begin{subequations}
\label{Pslp}
\begin{align}
\max\limits_{ \mathbf{x}_{m},t_{m}} & \quad t_{m}\\ \vspace{1ex}
 \begin{split}\text { s.t. } &~~~\big[ \Re\{\mathbf{h}_{k}^{H} \mathbf{x}_{m} e^{-j\angle{s_{m,k}}}\}- t_m\big] \tan \theta\\ \vspace{1ex}
                &\quad\quad \quad \quad -\big| \Im\{\mathbf{h}_{k}^{H} \mathbf{x}_m e^{-j\angle{s_{m,k}}}\}\big| \geq 0,  \quad \forall k,\end{split} \\\vspace{1ex}
                & \quad\|\mathbf{x}_m\|_2^2 \leq P_m,
\end{align}
\end{subequations}
where $t_{m}=\sigma\sqrt{\Gamma_{m}}$ denotes the communication quality-of-service (QoS) with $\Gamma_m$ representing the SNR for the $m$-th kind of symbol vector, and $P_m$ is the instantaneous transmit power budget for transmitting $\mathbf{x}_m$.

We observe that problem (\ref{Pslp}) is a second-order cone programming (SOCP) problem, which can be solved by using some optimization tools such as CVX \cite{CVX solver} or the efficient algorithm developed in \cite{A Li CF}.
However, since we need to solve this type of problem for $N_{\text{SLP}}=\Omega^K$ times, the overall complexity is unaffordable when the number of users is large.
In order to tackle this issue, in the following sections we first propose a novel G-SLP strategy to significantly reduce the number of precoders to be designed, and then develop efficient algorithms for obtaining each precoder.

\section{Concept of Grouped Symbol-Level Precoding and Problem Formulation}\label{sec: G-SLP}

\subsection{Grouped Symbol-Level Precoding} \label{sec:G-SLP A}
\begin{table}[t]
\caption{\label{tab1}Number of Designed Precoders for QPSK Modulation}
\centering

\begin{tabular}{p{1.6cm} p{1cm} p{1cm}  p{1cm} p{1cm} p{1cm}}
\toprule
{\diagbox{$N$}{$K$}} & 6 & 12  & 18 & 24 \\
\hline
$N_{\text{SLP}}$  & $4096$ & $1.68e7$  & $6.87e10$& $2.81e14$   \\
$N_{\text{G-SLP, 2 groups}}$  & $128$ & $8192$  & $5.24e5$ & $3.36e7$   \\
$N_{\text{G-SLP, 3 groups}}$ & $48$ & $768$  & $1.23e4$ & $1.97e5$   \\
\bottomrule
\end{tabular}\vspace{-2ex}
\end{table}

As described in Sec. II-B, the symbol information of all users are utilized to convert the MUI into CI for enhancing the detection performance.
Therefore, traditional SLP designs require the precoder optimizations for each different symbol vector of all users.
Consequently, the number of precoders to be designed is exponential with the user population, which is the major factor causing the high computational complexity in traditional SLP designs.
In an effort to provide low-complexity designs, we propose a novel G-SLP strategy to significantly lower the number of required optimizations.
In particular, we first divide all users into several groups and decompose the precoder designs for all users into the designs for the users in each group.
Instead of utilizing the symbol information of all users to fully exploit MUI, the precoder designs for the users in each group only utilize the symbol information of the users within this group to exploit the intra-group MUI as well as to suppress the interference caused to the users in other groups.
In the sequel, the number of precoders to be designed is exponential with the number of users in this group.
Therefore, by applying this G-SLP strategy, the total number of precoders to be designed is greatly reduced compared with that for conventional SLP scheme.

Specifically, we assume that the $K$ users are divided into $G$ groups by an appropriate grouping strategy as described in the next subsection.
The $g$-th group has $K_g$ users, $\sum_{g=1}^G K_g=K$.
Let $k_g$ denote the $k_g$-th user in the $g$-th group, $k_g = 1,\ldots,K_g$.
The transmitted symbol vector for the $g$-th group is denoted as $\mathbf{s}_{g,m_g}\triangleq[s_{g,m_g,1},\ldots,s_{g,m_g,K_g}]^T$, where $m_g$ denotes the $m_g$-th kind of symbol combination of the $g$-th group, $m_g=1,2,\ldots,\Omega^{K_g}$.
In the proposed G-SLP strategy, the transmitted signal that carries the required information symbols for all users is decomposed into $G$ precoded signals, each of which transfers the information symbols only for the users in a certain group.
In specific, for the transmitted vector $\mathbf{s}_m$ to the $K$ users, there exists a corresponding combination of symbol vectors $\{\mathbf{s}_{1,m_1},\ldots,\mathbf{s}_{G,m_G}\}$ of the $G$ groups.
Correspondingly, let $\mathbf{x}_{g,m_g} \in \mathbb{C}^{N_{\text{t}}}$ denote the precoder for transmitting $\mathbf{s}_{g,m_g}$ to the users in the $g$-th group. Thus, the precoder $\mathbf{x}_m$ for transmitting $\mathbf{s}_m$ can be constructed as the sum of the precoders $\{\mathbf{x}_{1,m_1},\ldots,\mathbf{x}_{G,m_G}\}$, i.e.,
\begin{equation}\label{Construction method}
\mathbf{x}_m=\sum_{g=1}^{G}\mathbf{x}_{g,m_g}.
\end{equation}

From (\ref{Construction method}), we can see that the $N_\text{SLP} = \Omega^K$ different precoders $\mathbf{x}_m$ are composed of $N_{\text{G-SLP}} = \sum_{g=1}^G \Omega^{K_g}$ different precoders $\mathbf{x}_{g,m_g}$ by applying the proposed G-SLP strategy.
Thanks to this ``intra-group symbol-level'' strategy, the number of precoders required to be designed is reduced to an exponential value of the number of users in each group.
In order to intuitively illustrate the advantages of the proposed strategy in decreasing the number of optimizations, we assume that the $K$ users are equally divided into $G$ groups and present some examples in Table \ref{tab1}, where $N_{\text{G-SLP, 2 groups}}$ and $N_{\text{G-SLP, 3 groups}}$ denote the total number of precoders to be designed when $G = 2$ and $G = 3$, respectively.
We observe that the proposed G-SLP strategy provides several orders of magnitude deduction in the number of precoders to be designed.
Moreover, it is obvious that the scheme with more groups requires much less number of optimizations.
However, considering that more users in each group may offer more MUI to be exploited for enhancing the performance, the number of groups $G$ is generally not very large.

With the G-SLP transmitted signal (\ref{Construction method}), when transmitting the $m$-th symbol combination $\mathbf{s}_m$, the received signal of the $k_g$-th user in the $g$-th group can be written as
\begin{equation}
\label{system model 2}
y_{m,g,k_g}=\underbrace{\mathbf{h}_{g,k_g}^H\mathbf{x}_{g,m_g}}_{ \substack{\text{desired signal with}\\ \text{ intra-group MUI}}} +\underbrace{\mathbf{h}_{g,k_g}^H\sum_{j=1,j\neq g}^G\mathbf{x}_{j,m_j} }\limits_{\text{inter-group\ MUI}} + n_{g,k_g},\end{equation}
where $\mathbf{h}_{g,k_g}$ denotes the channel vector of the $k_g$-th user in the $g$-th group, and $n_{g,k_g} \backsim\mathcal{CN}(0,\sigma^2)$ is the AWGN at the $k_g$-th user.
The received signal (\ref{system model 2}) consists of three parts: The desired signal with intra-group MUI, the inter-group MUI, and the AWGN.
Therefore, for achieving better symbol detection performance, the precoder designs should consider both the exploitation of intra-group MUI and the suppression of inter-group MUI.
In order to realize intra-group MUI exploitation, we follow the idea of SLP to design $\mathbf{x}_{g,m_g}$.
Specifically, the desired signal with intra-group MUI in (\ref{system model 2}), i.e., $\mathbf{h}_{g,k_g}^H\mathbf{x}_{g,m_g}$, is designed to be located in the constructive (green) region according to the symbol information $\mathbf{s}_{g,m_g}$.
We denote $\Gamma_{g,m_g}$ as the SNR requirement of the users in the $g$-th group for transmitting the $m_g$-th symbol combination $\mathbf{s}_{g,m_g}$, and use $t_{g,m_g}=\sigma\sqrt{ \Gamma_{g,m_g}}$ to equivalently represent the QoS requirement.
Then, similar to (\ref{mathmatical relation}), the QoS constraint for designing $\mathbf{x}_{g,m_g}$ is expressed as
\begin{equation}
\begin{aligned}
\label{mathmatical relation SLPG}
&\big[\Re\{{\mathbf{h}^H_{g,k_g}}\mathbf{x}_{g,m_g} e^{-j\angle{s_{g,m_g,k_g}}}\}-  t_{g,m_g} \big]\tan \theta\\
&\hspace{1 cm}-\big|\Im\{{\mathbf{h}^H_{g,k_g}}\mathbf{x}_{g,m_g} e^{-j\angle{s_{g,m_g,k_g}}}\}\big| \geq 0, \quad\forall k_g.
\end{aligned}
\end{equation}


On the other hand, according to (\ref{system model 2}), the precoders of one group inevitably cause interference to the users in other groups.
Therefore, it is necessary to suppress the inter-group MUI caused by $\mathbf{x}_{g,m_g}$ using the following constraint
\be
\label{intergroup interference condition}
    | \mathbf{h}_{j,k_j}^H\mathbf{x}_{g,m_g} |^2 \leq I_{g,m_g}, \quad \forall j\neq g,~\forall k_j ,\\
\ee
where the scalar $I_{g,m_g}$ denotes the allowed maximum inter-group MUI caused by $\mathbf{x}_{g,m_g}$.

Based on the above discussions, the max-min fairness problem for designing $\mathbf{x}_{g,m_g}$ can be formulated as
\begin{subequations}
\label{Pslpg}
\begin{align}
&\max\limits_{ \mathbf{x}_{g,m_g},t_{g,m_g}}  ~~ t_{g,m_g}\\ \label{constraint SLP}
\begin{split}&~~\text{ s.t. }~\big[ \Re\{\mathbf{h}_{g,k_g}^{H} \mathbf{x}_{g,m_g} e^{-j\angle{s_{g,m_g,k_g}}}\}-t_{g,m_g} \big] \tan \theta\\ \vspace{1ex}
&\quad\quad\quad~~-\big| \Im\{\mathbf{h}_{g,k_g}^{H} \mathbf{x}_{g,m_g} e^{-j\angle{s_{g,m_g,k_g}}}\}\big| \geq 0,  \quad\forall k_g,\end{split} \\
\label{interference suppress} &\quad\quad\quad| \mathbf{h}_{j,k_j}^{H} \mathbf{x}_{g,m_g}|^2 \leq I_{g,m_g}, \quad \forall j\neq g,~\forall k_j,\\
&\quad\quad\quad\|\mathbf{x}_{g,m_g}\|_2^2  \leq P_{g,m_g},
\end{align}
\end{subequations}
\nid where $P_{g,m_g}$ is the preset maximum transmit power for $\mathbf{x}_{g,m_g}$.
It is noted that each $\mathbf{x}_{g,m_g}$ is optimized to exploit the intra-group MUI by (\ref{Pslpg}b) as well as suppress the interference caused to the users in other groups by (\ref{Pslpg}c).
In addition, compared with the joint optimization, the precoders $\mathbf{x}_{g,m_g},~\forall g, m_g$, can be separately designed with much smaller dimensional variables and less number of constraints by introducing the axillary variables $I_{g,m_g}$ and $P_{g,m_g}$.
This method also enables more efficient solutions and the usage of powerful parallel computing to further accelerate the speed of optimizations.
Before solving the precoder design problem (\ref{Pslpg}), we observe that the grouping strategy, the power allocation, and the inter-group interference suppression are important issues that greatly influence the system performance.
Therefore, in the following subsections we will present the grouping strategy and the designs of $I_{g,m_g}$ and $P_{g,m_g}$.

\subsection{Grouping Strategy} \label{grouping strategy}

In order to achieve better system performance, the grouping strategy for the proposed G-SLP scheme aims to boost the intra-group MUI exploitation and suppress the harmful inter-group MUI.
Considering that the MUI of an MU-MISO system heavily depends on the correlation between the channels of the users, i.e., higher channel correlation generally causes stronger MUI and vice versa, we prefer relatively higher channel correlations within each group and lower channel correlations between them.

We use two metrics for measuring the channel correlations: One is to assess the correlation of two vectors $\mathbf{u}$ and $\mathbf{v}$: $
\rho(\mathbf{u},\mathbf{v})= \frac{|\mathbf{u}^H\mathbf{v}|}{\|\mathbf{u}\|_2\|\mathbf{v}\|_2}$;
the other one is to measure the overall correlation between a vector $\mathbf{u} \in \mathbb{C}^{n \times 1}$ and a matrix $\mathbf{Z}\in\mathbb{C}^{n\times m}$: $r(\mathbf{u},\mathbf{Z})=\frac{ \| \mathbf{P}_{\mathbf{Z}}\mathbf{u}\|_2 }{\|\mathbf{u}\|_2}$, where $\mathbf{P}_{\mathbf{Z}} \in\mathbb{C}^{n\times n}$ is the orthogonal projector onto $\mathbf{Z}$ and expressed as
\begin{equation}
\mathbf{P}_{\mathbf{Z}} =
\begin{cases}
\mathbf{Z}(\mathbf{Z}^H\mathbf{Z})^{-1}\mathbf{Z}^H , ~~\mathrm{if} ~~\mathrm{rank}(\mathbf{Z})=m,\\
\mathbf{B}(\mathbf{B}^H\mathbf{B})^{-1}\mathbf{B}^H , ~\mathrm{if} ~~\mathrm{rank}(\mathbf{Z})<m,
\end{cases}
\end{equation}
with $\mathbf{B}$ denoting the matrix whose columns are the basis for the vector space composed of $\mathbf{Z}\mathbf{x}$, $\forall \mathbf{x}\in \mathbb{C}^{m\times 1}$ \cite{Matrix book}.
It is noted that the metric $r(\mathbf{u},\mathbf{Z})$ is defined as the linear combination of correlations between $\mathbf{u}$ and each column of $\mathbf{Z}$ with normalization.
Then, by exploiting the channel correlation characteristics of the intra-group and inter-group users, we divide all users into several groups.
The detailed grouping procedure is described as follows.
\begin{itemize}
\item  Step 1.  Initialization,
\begin{align}
&\mathcal{K}=\{1,2,\ldots,K\},\\
&\mathcal{K}_g=\varnothing \,\text{(empty set)},\quad \forall g.
\end{align}
\item Step 2. Select the user who has the strongest channel condition as the first user of the first group,
\begin{equation}
\begin{aligned}
&\quad k_1^\star=\arg\max\limits_{k\in \mathcal{K}}~~\| \mathbf{h}_k \|_2, \\
&\mathcal{K}_1\leftarrow \mathcal{K}_1\cup \{k_1^\star\},\quad\mathcal{K} \leftarrow \mathcal{K}/\{k_1^\star\}.
\end{aligned}
\end{equation}
\item Step 3. Successively select the user who has the weakest inter-group channel correlation as the first user of other groups,
\begin{align}
&\quad k_g^\star=\arg\min\limits_{k\in \mathcal{K}}~\sum_{j\in \mathcal{K}_g,\forall g} \rho(\mathbf{h}_{j},\mathbf{h}_k), \label{sel1} \\
&\quad \mathcal{K}_g \leftarrow \mathcal{K}_g\cup \{k_g^\star\},\quad\mathcal{K} \leftarrow \mathcal{K}/\{k_g^\star\},
\end{align}
where $k_g^\star$ denotes the index of the newly selected user for the $g$-th group.
\item Step 4. Assign the remaining users to each group. For the $g$-th group, the newly joined user has the strongest channel correlation to the selected users, i.e., the $k_g^\star$-th user is selected by \vspace{-0.5ex}
    \begin{align}
    &k_g^\star=\arg\max\limits_{k\in \mathcal{K}}~~ r(\mathbf{h}_k,\mathbf{H}_{g,\text{r}}),\label{sel2}\\
    &\mathcal{K}_g \leftarrow \mathcal{K}_g\cup \{k_g^\star\},\quad\mathcal{K} \leftarrow \mathcal{K}/\{k_g^\star\},
    \end{align}
    where $\mathbf{H}_{g,\text{r}}$ is the channel matrix that contains the channel vectors of the selected users in the $g$-th group.
To balance the intra-group MUI and the inter-group MUI, each group successively selects its user according to (\ref{sel2}), until all users are assigned.
\end{itemize}

\vspace{-1.5ex}
\subsection{The Designs of $I_{g,m_{g}}$ and $P_{g,m_{g}}$} \label{set I P}

For the considered max-min fairness problem (\ref{Pslpg}), the power allocation to each precoder and the inter-group MUI tolerance will directly influence the QoS performance.
As introduced in Sec. III-A, the precoder designs for the users in each group should suppress the inter-group MUI caused to the users in other groups, which is similar to the idea of BLP.
Therefore, in this subsection we propose to pre-design the transmit power budget $P_{g,m_g}$ and the interference tolerance $I_{g,m_g}$ by leveraging the results of BLP designs, which can offer good power allocation and interference management in a cost-effective way.

In specific, for the considered max-min fairness case, we first obtain the block-level precoding matrix $\mathbf{W}\in \mathbb{C}^{N_{\text{t}}\times K}$ by maximizing the minimum SINR as well as satisfying the average transmit power budget $P_0$.
In particular, this classical BLP based SINR balancing problem is formulated as \vspace{-1.5ex}
\begin{subequations}
\begin{align}
\max\limits_{\mathbf{W},\gamma_0 }~~ &\gamma_0  \\
\text{s.t.} ~~~&\frac{|\mathbf{h}_k^H\mathbf{w}_k|^2}{  \sum_{j=1,j\neq k}^K|\mathbf{h}_k^H\mathbf{w}_j|^2 +\sigma^2   }\geq \gamma_0,k=1,\ldots,K,\\
&\|\mathbf{W}\|_F^2\leq P_0,
\end{align}
\end{subequations}
which can be efficiently solved using the iterative fixed-point algorithm proposed in \cite{optimal BLP}.
Then, the corresponding precoding matrix $\mathbf{W}_g \in \mathbb{C}^{N_{\text{t}}\times K_g}$ for each group can be extracted with the grouping result in the previous subsection.
Finally, the power budget $P_{g,m_g}$ for transmitting $\mathbf{x}_{g,m_g}$ is set as the power required to transfer symbol vector $\mathbf{s}_{g,m_g}$ via the BLP $\mathbf{W}_g$, i.e.,
\be\label{set of Pmg}
P_{g,m_g}=  \| \mathbf{W}_{g} \mathbf{s}_{g,m_g} \|_{2}^2.
\ee
With the power allocation strategy in (\ref{set of Pmg}), the average transmit power of the proposed G-SLP strategy can be calculated as
\begin{subequations}\label{eq:Pave}
\begin{align}
&P_{\text{ave}}\overset{(a)}{=} \mathbb{E} \Big\{\big\|\sum_{g=1}^G\mathbf{x}_{g,m_g}\big\|_2^2 \Big\}\hspace{-0.05cm}=\hspace{-0.05cm}\mathbb{E} \Big\{ \sum_{i=1}^G \sum_{j=1}^G \mathbf{x}_{i,m_i}^H \mathbf{x}_{j,m_j} \Big\}\\
& \overset{(b)}{=}\sum _{g=1}^G \mathbb{E}\big\{ \|\mathbf{x}_{g,m_g}\|_2^2 \big\}\hspace{-0.07cm}+ \hspace{-0.07cm}\sum_{i=1}^G\sum_{j=1,j\neq i}^G \hspace{-0.1cm}\mathbb{E}\{ \mathbf{x}_{i,m_i}^H\}\mathbb{E}\{ \mathbf{x}_{j,m_j}\}\\
&\overset{(c)}{=}\sum _{g=1}^G \mathbb{E} \big\{ \|\mathbf{x}_{g,m_g}\|_2^2 \big\} + 0\\
&\overset{(d)}{=} \sum_{g=1}^G\frac{1}{\Omega^{K_g}}\sum_{m_g=1}^{\Omega^{K_g}}\|\mathbf{x}_{g,m_g}\|_2^2 \\
& \overset{(e)}{\leq} \sum_{g=1}^G\frac{1}{\Omega^{K_g}}\sum_{m_g=1}^{\Omega^{K_g}} P_{g,m_g} \\
&\overset{(f)}{=} \sum_{g=1}^G\frac{1}{\Omega^{K_g}}\sum_{m_g=1}^{\Omega^{K_g}}\| \mathbf{W}_{g} \mathbf{s}_{g,m_g} \|_2^2\\
& \overset{(g)}{=} \|\mathbf{W}\|_F^2\leq P_0,
\end{align}
\end{subequations}
where (b) holds since the precoders of each group are independent, (c) holds due to the symmetry characteristic of the PSK symbols which results in $\mathbb{E} \{ \mathbf{x}_{g,m_g} \}=\mathbf{0},~\forall g$, (e) holds due to the power constraint (\ref{Pslpg}d), and (f) holds by (\ref{set of Pmg}).
From (\ref{eq:Pave}), we observe that the power allocation strategy (\ref{set of Pmg}) guarantees the average transmit power budget $P_0$.

Then, the inter-group MUI tolerance $I_{g,m_g}$ is chosen as the average interference from the $g$-th group to the users in other groups when transmitting $\mathbf{s}_{g,m_g}$, i.e.,
\be
I_{g,m_g}=\frac{1}{K-K_g}\sum_{\forall j\neq g, \forall k_j} \big| \mathbf{h}_{j,k_j}^H\mathbf{W}_{g}\mathbf{ s}_{g,m_g}\big|^2.\label{Img}
\ee

\section{Low-complexity Algorithm for\\  Max-Min Fairness Problem}
\label{sec: algorithm development for max-min}

In the previous section, we introduce a novel G-SLP strategy to significantly lower the required number of precoders to be designed.
For the max-min fairness problem (\ref{Pslpg}),  in this section we propose an efficient G-SLP design algorithm to further reduce the complexity.
With a linear approximation for the inter-group MUI suppression constraint, we simplify the design problem through Lagrangian and KKT conditions, then seek for a tractable surrogate objective function to utilize the MM method, and finally split the resulting problem into two sub-problems and derive the closed-form solutions to them.

\subsection{Problem Transformation}\label{problem transformation max-min}

We observe that the max-min fairness problem (\ref{Pslpg}) is a convex problem with ${K_g}$ linear constraints and $(K-K_g + 1)$ quadratic constraints.
Although it can be solved by applying some existing convex optimization tools such as the CVX toolbox, the computational complexity is prohibitive due to the numerous quadratic constraints.
In an effort to provide an efficient solution, in this subsection we propose to transform problem (\ref{Pslpg}) into a simpler one with only linear constraints.

\begin{figure}[!t]
\centering
\includegraphics[width = 2.6 in]{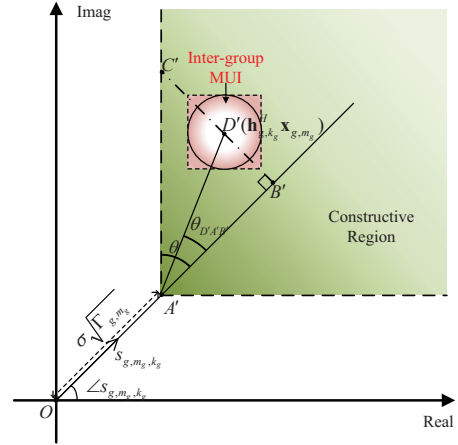}
\caption{Constructive region for a QPSK symbol.}\label{fig2} \vspace{-4 ex}
\end{figure}

Firstly, it is noted that the inter-group MUI should be suppressed within a certain level through the quadratic constraints (\ref{interference suppress}), by which the inter-group MUI is restricted to a red circle area around point $D'$, as shown in Fig. 2.
In order to efficiently obtain closed-form solutions, we relax (\ref{interference suppress}) into the following simpler linear constraints without compromising the purpose of suppressing the inter-group MUI:
\begin{subequations}
\label{new constraints}
\begin{align}
&   | \Re\{ \mathbf{h}_{j,k_j}^{H} \mathbf{x}_{g,m_g}\}| \leq \sqrt{ I_{g,m_g} }, \quad   \forall j\neq g,~\forall k_j,\\
&   | \Im\{ \mathbf{h}_{j,k_j}^{H} \mathbf{x}_{g,m_g}\}| \leq \sqrt{ I_{g,m_g} }, \quad   \forall j\neq g,~\forall k_j,
\end{align}
\end{subequations}
by which the inter-group MUI is limited to a squared region that has the same center and radius with the circle area endowed by constraints (\ref{interference suppress}).
Since the feasible region maintains almost the same after relaxation, the resulting performance loss will be very marginal.
Then, to facilitate the algorithm development, we equivalently re-arrange the QoS constraints (\ref{constraint SLP}) and the interference suppression constraints (\ref{new constraints}) by decomposing each absolute operation into two separate operations, and re-formulate the relaxation of the max-min fairness problem (\ref{Pslpg}) as
\begin{subequations}
\label{Pslpgs}
\begin{align}
&\max\limits_{ \mathbf{x}_{g,m_g},t_{g,m_g}}\quad t_{g,m_g}\\
\begin{split}&~~~\text { s.t. }~[ \Re\{\mathbf{h}_{g,k_g}^{H} \mathbf{x}_{g,m_g} e^{-j\angle{s_{g,m_g,k_g}}}\}- t_{g,m_g} ] \tan \theta\\
& \quad\quad\quad\quad~\pm\Im\{\mathbf{h}_{g,k_g}^{H} \mathbf{x}_{g,m_g} e^{-j\angle{s_{g,m_g,k_g}}}\} \geq 0,  ~~\forall k_g, \end{split} \\
&\quad\quad\quad\pm  \Re\{ \mathbf{h}_{j,k_j}^{H} \mathbf{x}_{g,m_g}\} \leq \sqrt{ I_{g,m_g} }, \quad   \forall j\neq g,~\forall k_j,\\
&\quad\quad\quad\pm \Im\{ \mathbf{h}_{j,k_j}^{H} \mathbf{x}_{g,m_g}\} \leq \sqrt{ I_{g,m_g} }, \quad   \forall j\neq g,~\forall k_j,
\label{linear relaxation}\\
&\quad\quad\quad\|\mathbf{x}_{g,m_g}\|_2^2  \leq P_{g,m_g}.
\end{align}
\end{subequations}
In order to efficiently deal with this complex-valued problem, we further transform it into a concise real-valued problem by defining \vspace{-1ex}
\begin{subequations}
\label{complex to real}
\begin{align}
\widetilde{\mathbf{x}}_{g,m_g} &\triangleq {[\Re\{\mathbf{x}_{g,m_g}\}^T,\Im\{\mathbf{x}_{g,m_g}\}^T] }^T, \\
\begin{split}
\widetilde{\mathbf{H}}_{g,m_g}&\triangleq \left[ \begin{aligned}
&\Re\{\mathbf{H}_{g} \text{diag}( e^{ j\angle{ \mathbf{s}_{g,m_g} } } ) \} \\
&\Im\{\mathbf{H}_{g} \text{diag}( e^{ j\angle{ \mathbf{s}_{g,m_g} } }) \}\end{aligned} \right],\end{split}\\
\mathbf{c}_1&\triangleq\tan \theta  [\mathbf{1}_{2K_g \times 1}^T,\mathbf{0}_{4(K-K_g) \times 1}^T]^T , \label{c1}\\
\mathbf{c}_2&\triangleq [\mathbf{0}_{2K_g \times 1}^T,\mathbf{1}_{4(K-K_g) \times 1}^T]^T,\\
\begin{split}
\boldsymbol{\nabla} &\triangleq\left[\begin{array}{lr}
\mathbf{0}_{N_{\text{t}} \times N_{\text{t}}} & \mathbf{I}_{N_{\text{t}} \times N_{\text{t}}} \\
-\mathbf{I}_{N_{\text{t}} \times N_{\text{t}}} & \mathbf{0}_{N_{\text{t}} \times N_{\text{t}}}
\end{array}\right],
\end{split}\\
\widetilde{\mathbf{H}}_{g}^{\text{c}} &\triangleq {[\Re\{\mathbf{H}_{g}^{\text{c}}\}^T,\Im\{\mathbf{H}_{g}^{\text{c}}\}^T]}^T,\\
\label{Hc}
\mathbf{A}_{g,m_g} &\triangleq \left[\hspace{-0.4cm}\begin{array}{lc}
&\,\,\,\,\,\widetilde{\mathbf{H}}_{g,m_g}^T\boldsymbol{\nabla}  -\tan\theta  \widetilde{\mathbf{H}}_{g,m_g}^T\\
&-          \widetilde{\mathbf{H}}_{g,m_g}^T\boldsymbol{\nabla} -\tan\theta  \widetilde{\mathbf{H}}_{g,m_g}^T\\
&\,\,\,\,\,({ \widetilde{\mathbf{H}}_{g}^{\text{c}} })^T \\
&-          ({\widetilde{\mathbf{H}}_{g}^{\text{c}}})^T  \\
&\,\,\,\,\,({\widetilde{\mathbf{H}}_{g}^{\text{c}}})^T \boldsymbol{\nabla}\\
&-          ({\widetilde{\mathbf{H}}_{g}^{\text{c}}})^T \boldsymbol{\nabla}
\end{array}\right],
\end{align}
\end{subequations}
where $\mathbf{H}_g\in \mathbb{C}^{N_{\text{t}}\times K_g }$ consists of the channel vectors of the users in the $g$-th group and the arrangement of them is in the same order as $\mathbf{s}_{g,m_g}$, and $\mathbf{H}_g^{\text{c}}\in \mathbb{C}^{N_{\text{t}}\times (K-K_g) }$ consists of all channel vectors of the users out of the $g$-th group.
Then, the equivalent compact real-valued form of problem (\ref{Pslpgs}) is expressed as \vspace{-1ex}
\begin{subequations}
\label{Pslpg compact}
\begin{align}
\min\limits_{ \widetilde{\mathbf{x}}_{g,m_g},t_{g,m_g}}& -t_{g,m_g}\\
\text { s.t. } \quad &  \mathbf{A}_{g,m_g}\widetilde{\mathbf{x}}_{g,m_g}+  t_{g,m_g}  \mathbf{c}_1 - \sqrt{ I_{g,m_g} }  \mathbf{c}_2 \leq 0, \label{linear constraints Pslpg compact}\\
                 &   \widetilde{\mathbf{x}}_{g,m_g}^T\widetilde{\mathbf{x}}_{g,m_g} - P_{g,m_g} \leq 0. \label{quadratic constraint Pslpg compact}
\end{align}
\end{subequations}

It is obvious that problem (\ref{Pslpg compact}) is convex and can be solved using the CVX toolbox.
However, its $(4K-2K_g)$ linear constraints (\ref{linear constraints Pslpg compact}) and one quadratic constraint (\ref{quadratic constraint Pslpg compact}) still bring high computational complexity. Fortunately, we observe that the Slater's condition is satisfied \cite{convex opt Boyd}.
Thus, solving the Lagrangian dual problem of (\ref{Pslpg compact}) with considering the KKT conditions can provide a low-complexity solution.
Particularly, its Lagrangian can be expressed as
\begin{align}
&\mathcal{L}_1( t_{g,m_g},\widetilde{\mathbf{x}}_{g,m_g},\boldsymbol{\mu},\mu_0)= -t_{g,m_g} +\mu_0(\widetilde{\mathbf{x}}_{g,m_g}^T\widetilde{\mathbf{x}}_{g,m_g} - P_{g,m_g}) \nonumber \\
&\quad+ \boldsymbol{\mu}^T(\mathbf{A}_{g,m_g}\widetilde{\mathbf{x}}_{g,m_g}+ t_{g,m_g} \mathbf{c}_1 - \sqrt{I_{g,m_g} }  \mathbf{c}_2),
\end{align}
where $\boldsymbol{\mu}\triangleq[\mu_1,\mu_2,\ldots,\mu_{4K-2K_g}]$ and $\mu_0$ are dual variables, $\mu_0\geq 0$, $\mu_i \geq 0,\forall i$.
The dual problem of (\ref{Pslpg compact}) is thus given by
\begin{align}
\label{Pdual}
\max\limits_{\boldsymbol{\mu},\mu_0}~&\min\limits_{\hspace{0.1cm} t_{g,m_g},\widetilde{\mathbf{x}}_{g,m_g}} \mathcal{L}_1( t_{g,m_g},\widetilde{\mathbf{x}}_{g,m_g},\boldsymbol{\mu},\mu_0) \nonumber\\
 &~\quad\quad\text{s.t.}~~\quad \mu_0 \geq 0,~\mu_i\geq 0, \quad \forall i.
\end{align}
According to the KKT conditions for optimality, the optimal solution to problem (\ref{Pslpg compact}) and its dual problem (\ref{Pdual}) should satisfy
\begin{subequations}
\label{KKT}
\begin{align} \vspace{1ex}
&\frac{\partial \mathcal{L}_1}{\partial t_{g,m_g} }=\boldsymbol{\mu}^T\mathbf{c}_1-1=0, \label{KKT1} \\
&\frac{\partial \mathcal{L}_1}{\partial \widetilde{\mathbf{x}}_{g,m_g}}=\mathbf{A}_{g,m_g}^T\boldsymbol{\mu}+2\mu_0\widetilde{\mathbf{x}}_{g,m_g} =\mathbf{0}, \label{KKT2}\\
&  \boldsymbol{\mu} \odot (\mathbf{A}_{g,m_g}\widetilde{\mathbf{x}}_{g,m_g}+ t_{g,m_g}  \mathbf{c}_1 - \sqrt{ { I_{g,m_g} }}  \mathbf{c}_2) = \mathbf{0}, \label{KKT3}\\
& \mu_0(\widetilde{\mathbf{x}}_{g,m_g}^T\widetilde{\mathbf{x}}_{g,m_g} - P_{g,m_g})=0, \label{KKT4}\\
& \mu_0\geq 0, \label{KKT5} \\
& \mu_i\geq 0,\quad \forall i. \label{KKT6}
\end{align}
\end{subequations}
Based on (\ref{KKT1}), (\ref{KKT2}) and (\ref{complex to real}), we have $\mu_0\neq 0$ and consequently the term in the parenthesis of (\ref{KKT4}) is $0$. Thus, the optimal $\mu_0^{\star}$ and $\widetilde{\mathbf{x}}_{g,m_g}^{\star}$ can be expressed as \vspace{-0.1cm}
\begin{subequations}
\label{x mu0}
\begin{align}
\mu_0^{\star}&=\sqrt{ {{\boldsymbol{\mu}^{\star}}^T\mathbf{A}_{g,m_g}\mathbf{A}_{g,m_g}^T\boldsymbol{\mu}^{\star}}/{(4P_{g,m_g}})},\\
\widetilde{\mathbf{x}}_{g,m_g}^{\star}&=-{\mathbf{A}_{g,m_g}^T\boldsymbol{\mu}^{\star}}/({2\mu_0^{\star}}),\label{xmao}
\end{align}
\end{subequations}
where $\boldsymbol{\mu}^{\star}$ is the optimal solution to the dual problem (\ref{Pdual}).
Applying the KKT condition (\ref{KKT1}) and the results in (\ref{x mu0}) into (\ref{Pdual}), we can further simplify the dual problem as \vspace{-0.1cm}
\begin{subequations}
\label{Pdual 2}
\begin{align} \vspace{1ex}
\label{dual 1}
&\max\limits_{\boldsymbol{\mu},\mu_0}~\min\limits_{ t_{g,m_g},\widetilde{\mathbf{x}}_{g,m_g}}~\mathcal{L}_1( t_{g,m_g},\widetilde{\mathbf{x}}_{g,m_g},\boldsymbol{\mu},\mu_0)\\
\label{dual 3}
&=\max\limits_{\boldsymbol{\mu},\mu_0}~ - \frac{ \boldsymbol{\mu}^T\mathbf{A}_{g,m_g}\mathbf{A}_{g,m_g}^T\boldsymbol{\mu} }{2\mu_0} - \sqrt{  I_{g,m_g} } \boldsymbol{\mu}^T \mathbf{c}_2\\
\label{dual 4}
&=\max\limits_{\boldsymbol{\mu}}  ~-\hspace{-0.05 cm}\sqrt{P_{g,m_g}\boldsymbol{\mu}^T \mathbf{A}_{g,m_g}\mathbf{A}_{g,m_g}^T \boldsymbol{\mu}}\hspace{-0.05 cm} -\hspace{-0.1 cm}\sqrt{  I_{g,m_g} } \boldsymbol{\mu}^T \mathbf{c}_2\\
&= \max\limits_{\boldsymbol{\mu}}  ~ -\sqrt{P_{g,m_g}}\big\|\mathbf{V}_{g,m_g}^{\frac{1}{2}}\boldsymbol{\mu}\big\|_2 - \sqrt{  I_{g,m_g} } \boldsymbol{\mu}^T \mathbf{c}_2,
\label{dual 5}
\end{align}
\end{subequations}
where we define \vspace{-0.1cm}
\begin{equation}
\label{get V}
\mathbf{V}_{g,m_g}\triangleq  \mathbf{ A}_{g,m_g}\mathbf{A}_{g,m_g}^T.
\end{equation}
Finally, with the constraints (\ref{KKT1}) and (\ref{KKT6}) on the dual variable $\boldsymbol{\mu}$, the Lagrangian dual problem (\ref{Pdual}) can be equivalently re-written as \vspace{-0.1cm}
\begin{subequations}
\label{Psimple}
\begin{align}
\min\limits_{\boldsymbol{\mu}}\quad &f(\boldsymbol{\mu}) \triangleq \sqrt{P_{g,m_g}}\|\mathbf{V}_{g,m_g}^{\frac{1}{2}}\boldsymbol{\mu}\|_2+\sqrt{  I_{g,m_g}} \boldsymbol{\mu}^T \mathbf{c}_2 \label{objective func}\\
\text{s.t.} \quad & \boldsymbol{\mu}^T\mathbf{c}_1 -1 =0,\\
            & \mu_i\geq 0, \quad \forall i.
\end{align}
\end{subequations}
Based on the above derivations, after solving problem (\ref{Psimple}) and obtaining the dual variable $\boldsymbol{\mu}^{\star}$, the optimal solution $\widetilde{\mathbf{x}}_{g,m_g}^{\star}$ to its original problem (\ref{Pslpg compact}) can be calculated by (\ref{x mu0}), and the complex-valued solution $\mathbf{x}_{g,m_g}^{\star}$ is constructed by \vspace{-0.1cm}
\begin{equation}
\label{get xstar}
\mathbf{x}_{g,m_g}^{\star}=\mathbf{U}\widetilde{\mathbf{x}}_{g,m_g}^{\star},
\end{equation}
where $\mathbf{U}\triangleq[ \mathbf{I}_{ N_{\text{t} } } \; j\mathbf{I}_{N_{ \text{t}}  }]$ is a transformation matrix that transforms the real-valued vector $\widetilde{\mathbf{x}}_{m,g}^{\star}$ into its complex equivalence.

We observe that the optimization problem (\ref{Psimple}) has simple linear constraints, but its objective function is not easy to be handled due to the 2-norm term.
In order to solve this problem in closed-form, in next subsection we first find a more tractable surrogate objective function to employ the MM method and then develop efficient algorithms for the resulting problem.

\subsection{MM-based Iterative Algorithm}
\label{sec:MM for max-min}

Since the 2-norm term in the objective (\ref{Psimple}a) prevents a direct solution, we propose to find a preferable upper-bounded surrogate function to locally approximate the objective function (\ref{Psimple}a) in each iteration by employing the MM method \cite{MM}.
Particularly, an upper-bound of $\|\mathbf{V}_{g,m_g}^{\frac{1}{2}}\boldsymbol{\mu}\|_2$ can be derived as \vspace{-0.1cm}
\begin{equation}
\label{upper bound1}
\big\|\mathbf{V}_{g,m_g}^{\frac{1}{2}}\boldsymbol{\mu}\big\|_2  \leq \frac{ \big\|\mathbf{V}_{g,m_g}^{\frac{1}{2}}\boldsymbol{\mu}\big\|_2^2 +\big\|\mathbf{V}_{g,m_g}^{\frac{1}{2}}\boldsymbol{\mu}_t\big\|_2^2 }{2\big\|\mathbf{V}_{g,m_g}^{\frac{1}{2}}\boldsymbol{\mu}_t\big\|_2},
\end{equation}
where $\boldsymbol{\mu}_t$ denotes the solution in the previous $t$-th iteration.
In an effort to find a more tractable surrogate objective function, we further derive an upper bound for the first term of the numerator in (\ref{upper bound1}) by exploiting the quadratic Taylor expansion and Rayleigh quotient as: \vspace{-0.1cm}
\begin{subequations}
\label{upper bound construct}
\begin{align}
\begin{split}
\big\| \mathbf{V}_{g,m_g}^{\frac{1}{2}}\boldsymbol{\mu}\big\|_2^2 =&\ \boldsymbol{\mu}_t^T \mathbf{V}_{g,m_g} \boldsymbol{\mu}_t+2\boldsymbol{\mu}_t^T\mathbf{V}_{g,m_g}^T(\boldsymbol{\mu}-\boldsymbol{\mu}_t)\\
 &\ +(\boldsymbol{\mu}-\boldsymbol{\mu}_t)^T\mathbf{V}_{g,m_g}(\boldsymbol{\mu}-\boldsymbol{\mu}_t)\end{split}\\
 \begin{split}
 \label{upper bound2}
\leq &\ \boldsymbol{\mu}_t^T \mathbf{V}_{g,m_g} \boldsymbol{\mu}_t+2\boldsymbol{\mu}_t^T\mathbf{V}_{g,m_g}^T(\boldsymbol{\mu}-\boldsymbol{\mu}_t)\\
&\ +\lambda_{g,m_g}(\boldsymbol{\mu}-\boldsymbol{\mu}_t)^T(\boldsymbol{\mu}-\boldsymbol{\mu}_t),\end{split}
\end{align}
\end{subequations}
where $\lambda_{g,m_g}$ denotes the maximum eigenvalue of $\mathbf{V}_{g,m_g}$.
Plugging the results of (\ref{upper bound construct}) and (\ref{upper bound1}) into (\ref{objective func}), an upper-bound surrogate function of $f(\bm{\mu})$ is formulated as \vspace{-0.1cm}
\begin{equation}
\label{surrogate function max-min}
\begin{aligned}
\widetilde{f}(\boldsymbol{\mu} |\boldsymbol{\mu}_t)\triangleq\alpha\boldsymbol{\mu}^T\boldsymbol{\mu}+\mathbf{q}^T\boldsymbol{\mu}+\alpha\boldsymbol{\mu}_t^T\boldsymbol{\mu}_t,
\end{aligned}
\end{equation}
where for simplicity we define \vspace{-0.1cm}
\begin{subequations}\label{alpha q}
\begin{align}
\alpha\triangleq&\frac{\sqrt{P_{g,m_g}}\lambda_{g,m_g}}{2\big\|\mathbf{V}_{g,m_g}^{\frac{1}{2}}\boldsymbol{\mu}_t\big\|_2},\\
\mathbf{q}\triangleq& \frac{\sqrt{P_{g,m_g}}(\mathbf{V}_{g,m_g}-\lambda_{g,m_g}\mathbf{I})\bm{\mu}_t}
{\big\|\mathbf{V}_{g,m_g}^{\frac{1}{2}}\boldsymbol{\mu}_t\big\|_2}+\sqrt{I_{g,m_g}}\mathbf{c}_2.
\end{align}
\end{subequations}
Therefore, the optimization problem in each iteration can be formulated as \vspace{-0.1cm}
\begin{subequations}
\label{P-minimization}
\begin{align}
\min\limits_{\boldsymbol{\mu}} &\quad \alpha\boldsymbol{\mu}^T\boldsymbol{\mu}+\mathbf{q}^T\boldsymbol{\mu}\\
\text{s.t.} & \quad\boldsymbol{\mu}^T\mathbf{c}_1 -1 =0, \label{equation constraint P-minimization}\\
            & \quad\mu_i\geq 0,\quad\forall i.
\end{align}
\end{subequations}

\vspace{-0.1cm}
We see that the objective function (\ref{P-minimization}a) is quadratic and separable in the variable $\bm{\mu}$, which allows problem (\ref{P-minimization}) to be addressed more easily.
Furthermore, we also observe that most elements of $\mathbf{c}_1$ are zeros as defined in (\ref{c1}).
Therefore, problem (\ref{P-minimization}) can be divided into two sub-problems, which are of lower dimensions and can be efficiently solved in parallel.
For brevity, we first define \vspace{-0.1cm}
\begin{equation}
\label{mu q decompose}
\boldsymbol{\mu}_t\triangleq[\boldsymbol{\mu}_{t,1} ^T,\boldsymbol{\mu}_{t,2}^T]^T ,\ \boldsymbol{\mu}\triangleq[\boldsymbol{\mu}_1^T, \boldsymbol{\mu}_2^T]^T,\ \mathbf{q}\triangleq[\mathbf{q}_1^T,\mathbf{q}_2^T]^T,
\end{equation}
where $\boldsymbol{\mu}_{t,1}$, $\boldsymbol{\mu}_1$, and $\mathbf{q}_1$ are the first to the $2K_g$-th elements of $\boldsymbol{\mu}_t$, $\boldsymbol{\mu}$, and $\mathbf{q}$, respectively, and $\boldsymbol{\mu}_{t,2}$, $\boldsymbol{\mu}_2$, and $\mathbf{q}_2$ are the remaining $(2K_g+1)$-th to $(4K-2K_g)$-th elements of $\boldsymbol{\mu}_t$, $\boldsymbol{\mu}$, and $\mathbf{q}$, respectively.
Then, problem (\ref{P-minimization}) can be equivalently decomposed into two sub-problems with respect to $\bm{\mu}_1$ and $\bm{\mu}_2$.
One is to solve for $\bm{\mu}_1$ as follows \vspace{-0.1cm}
\begin{subequations}
\label{P-minimization 1}
\begin{align} \label{objective func P-minimization 1}
\min\limits_{\boldsymbol{\mu}_1} & \quad \widetilde{f}(\boldsymbol{\mu}_1|\boldsymbol{\mu}_{t})\triangleq\alpha\boldsymbol{\mu}_1^T\boldsymbol{\mu}_1+\mathbf{q}_1^T\boldsymbol{\mu}_1\\
\text{s.t.} & \quad\tan\theta  \mathbf{1}^T\boldsymbol{\mu}_1 -1 =0, \label{equality constraint P-minimization 1}\\
            & \quad \mu_{\text{1},i}\geq 0,\quad \forall i,
\end{align}
\end{subequations}
where $\mu_{\text{1},i}$ denotes the $i$-th element of $\boldsymbol{\mu}_1$.
The other is to solve for $\boldsymbol{\mu}_2$ as follows
\begin{subequations}
\label{P-minimization 2}
\begin{align}\vspace{1ex}
\min\limits_{\boldsymbol{\mu}_2} & \quad \widetilde{f}(\boldsymbol{\mu}_2|\boldsymbol{\mu}_{t})\triangleq \alpha\boldsymbol{\mu}_2^T\boldsymbol{\mu}_2+\mathbf{q}_2^T\boldsymbol{\mu}_2\\ \vspace{1ex}
\text{s.t.} & \quad \mu_{\text{2},i}\geq 0,\quad \forall i, \label{non negative constraint P-minimization 2}
\end{align}
\end{subequations}
where $\mu_{\text{2},i}$ denotes the $i$-th element of $\boldsymbol{\mu}_2$.
In the follows, we present efficient solutions for problems (\ref{P-minimization 1}) and (\ref{P-minimization 2}), respectively.

Problem (\ref{P-minimization 1}) is a $2K_g$-dimensional quadratic programming (QP) problem with one linear equality constraint and $2K_g$ non-negative constraints.
Although it can be solved by some existing algorithms, e.g., interior-point method and active set method \cite{ASM}, we propose an iterative method by exploiting its Lagrangian dual to further alleviate the computing burden.
Specifically, the Lagrangian of problem (\ref{P-minimization 1}) is written as
\begin{equation}
\mathcal{L}_2(\lambda,\boldsymbol{\mu}_1|\boldsymbol{\mu}_{t}) = \alpha\boldsymbol{\mu}_1^T\boldsymbol{\mu}_1+\mathbf{q}_1^T\boldsymbol{\mu}_1+\lambda(\tan\theta  \mathbf{1}^T\boldsymbol{\mu}_1 -1),
\end{equation}
where $\lambda \in \mathbb{C}$ is a dual variable.
In this form, the Lagrange dual problem of (\ref{P-minimization 1}) is formulated as
\begin{subequations}
\label{P dual sovle mu1}
\begin{align}
\max\limits_{\lambda} \; \min\limits_{\boldsymbol{\mu}_1}& \;\;\mathcal{L}_2(\lambda,\boldsymbol{\mu}_1|\boldsymbol{\mu}_{t})\\
\text{s.t.} &\;\;\mu_{\text{1},i}\geq 0,\quad \forall i, \label{non-neg Pmu1}
\end{align}
\end{subequations}
which is a two-block problem and can be efficiently solved by alternatively updating the variables $\bm{\mu}_1$ and $\lambda$.
Given $\lambda$, by setting the partial derivation with respect to $\bm{\mu}_1$ to zero
\begin{equation}
\label{partial derivation1}
\begin{array}{rll}
\frac{\partial \mathcal{L}_2(\lambda,\boldsymbol{\mu}_1|\boldsymbol{\mu}_{t}) }{\partial \bm{\mu}_1 } =2\alpha\bm{\mu}_1+\mathbf{q}_{1}+\lambda\tan\theta \mathbf{1}=\mathbf{0},
\end{array}
\end{equation}
the conditionally optimal $\bm{\mu}_1^\star$ can be calculated as
\begin{equation}
\label{get mu1}
\bm{\mu}_1^{\star}=\max\Big\{\mathbf{0},-\frac{\lambda\tan\theta\mathbf{1}+ \mathbf{q}_{1}}{2\alpha}\Big\}.
\end{equation}
Then with $\bm{\mu}_1^\star$, the dual problem (\ref{P dual sovle mu1}) is reduced to a univariate problem with respect to the dual variable $\lambda$ as
\begin{equation}\begin{aligned}
\label{P dual sovle simple}
g(\lambda)&\triangleq \max\limits_{\lambda} \;\mathcal{L}_2(\lambda,\boldsymbol{\mu}_1^\star|\boldsymbol{\mu}_{t}),\\
&= \max\limits_{\lambda} \; \lambda(\tan\theta  \mathbf{1}^T\boldsymbol{\mu}_1 -1),
\end{aligned}\end{equation}
which can be easily solved by the gradient ascent method \cite{convex opt Boyd}.
Specifically, the update of $\lambda^{(j+1)}$ in the $(j+1)$-th iteration is given by
\be
\lambda^{(j+1)}=\lambda^{(j)}+ a^{(j+1)}\nabla \mathcal{L}_2(\lambda,\boldsymbol{\mu}_1^{\star}|\boldsymbol{\mu}_{t}),\label{dual ascent step2}
\ee
where $a^{(j+1)} > 0$ is the step size obtained by the line search method, and the gradient $\nabla \mathcal{L}_2(\lambda,\boldsymbol{\mu}_1^{\star}|\boldsymbol{\mu}_{t}) = \tan\theta  \mathbf{1}^T\boldsymbol{\mu}_1^\star -1$ according to (\ref{P dual sovle simple}).
The above mentioned algorithm to solve problem (\ref{P-minimization 1}) is summarized in Algorithm \ref{subAlgorithm for max-min}, where $\epsilon_1$ is a parameter to judge the convergence.

\begin{algorithm}[t]\begin{small}
  \caption{Efficient Algorithm to Solve Problem (\ref{P-minimization 1})}
  \label{subAlgorithm for max-min}
  \begin{algorithmic}[1]
    \REQUIRE $\alpha,\mathbf{q}_{1}$, $\epsilon_1$.
    \ENSURE  $\boldsymbol{\mu}^\star_{1}$.
    \STATE {Initialize $j := 0$, $\lambda^{(0)}$, $a^{(0)}$.}
    \REPEAT
    \STATE {Update $\boldsymbol{\mu}_{1}^{(j+1)}$ by (\ref{get mu1});}
    \STATE {Calculate the step size $a^{(j+1)}$ by the line search method;}
    \STATE {Update $\lambda^{(j+1)}$ by (\ref{dual ascent step2});}
    \STATE {$j := j + 1$;}
    \UNTIL {$\nabla \mathcal{L}_2(\lambda,\boldsymbol{\mu}_1^{(j)}|\boldsymbol{\mu}_{t}) \leq \epsilon_1$.}
    \STATE {$\bm{\mu}_1^\star := \bm{\mu}_1^{(j)}$.}
  \end{algorithmic}\end{small}
\end{algorithm}

Problem (\ref{P-minimization 2}) has the same form as the inner minimization problem of (\ref{P dual sovle mu1}).
Therefore, the optimal solution $\bm{\mu}_2^\star$ to problem (\ref{P-minimization 2}) can be easily calculated by
\begin{equation}
\label{get mu2}
\bm{\mu}_2^{\star}=\max\Big\{\mathbf{0},-\frac{\mathbf{q}_2}{2\alpha}\Big\}.
\end{equation}

\subsection{Algorithm Summary and Complexity Analysis}\label{sec: complexity analysis for max-min}

Based on the above derivations, the proposed low-complexity G-SLP design algorithm for the max-min fairness problem is straightforward and summarized in Algorithm \ref{Algorithm for max-min}, where $\epsilon_2$ is a parameter to judge the convergence.
We iteratively update the dual variable $\bm{\mu}$, which is split into two parts: One is solved by Algorithm 1 and the other is calculated in a closed-form (\ref{get mu2}), then obtain the solution to the original real-valued problem (\ref{Pslpgs}) according to the KKT conditions (\ref{x mu0}), and finally construct the complex-valued solution by (\ref{get xstar}).

\begin{algorithm}[t]\begin{small}
  \caption{MM-based G-SLP Design Algorithm for Max-Min Fairness Problem (\ref{Pslpgs})}
  \label{Algorithm for max-min}
  \begin{algorithmic}[1]
    \REQUIRE $\mathbf{H}_g$, $\mathbf{s}_{g,m_g}$, $I_{g,m_g}$, $P_{g,m_g}$,  $\epsilon_2$.
    \ENSURE  $\mathbf{x}_{g,m_g}^{\star}$.
    \STATE {Initialize $t := 0$, $\boldsymbol{\mu}_{0}:=[\mathbf{c}_1^T/\|\mathbf{c}_1\|_1,\mathbf{0}^T]^T$.}
    \STATE {Calculate $\mathbf{A}_{g,m_g}$ by (\ref{Hc}), $\mathbf{V}_{g,m_g}$ by (\ref{get V}), and $\lambda_{g,m_g}$.}
    \REPEAT
    \STATE {Calculate $\alpha$ by (\ref{alpha q}a), $\mathbf{q}$ by (\ref{alpha q}b), $\mathbf{q}_{1}$ and  $\mathbf{q}_{2}$ by (\ref{mu q decompose});}
    \STATE {Update $\boldsymbol{\mu}_{t+1,1}$ by solving (\ref{P dual sovle mu1}) via Algorithm \ref{subAlgorithm for max-min};}
    \STATE {Update $\boldsymbol{\mu}_{t+1,2}$ by (\ref{get mu2});}
    \STATE {$t := t + 1$;}
    \UNTIL {$\|\boldsymbol{\mu}_{t} - \boldsymbol{\mu}_{t-1}\|_2/\| \boldsymbol{\mu}_{t-1}\|_2\leq \epsilon_2$ }.
    \STATE{$\bm{\mu}^\star := \boldsymbol{\mu}_{t}$.}
    \STATE {Calculate $\widetilde{\mathbf{x}}_{g,m_g}^{\star}$ by (\ref{x mu0});}
    \STATE {Construct $\mathbf{x}_{g,m_g}^{\star}$ by (\ref{get xstar}).}
  \end{algorithmic}\end{small}
\end{algorithm} 

The computational complexity of steps 1-2 in Algorithm 2 depends on the calculation of $\lambda_{g,m_g}$, which has complexity of order $\mathcal{O}\big( \left(4K-2K_g\right)^3 \big)$.
The complexity to calculate $\alpha$ and $\mathbf{q}$ is of order $\mathcal{O}\big( (4K-2K_g)^2 \big)$.
Updating $\bm{\mu}_{t+1,1}$ by Algorithm 1 has complexity of order $\mathcal{O}\left(6N_\mathrm{in}K_g \right)$, where $N_\mathrm{in}$ represents the number of iterations in the inner loop, and updating $\boldsymbol{\mu}_{t+1,2}$ has complexity of order $\mathcal{O}\left(4K-4K_g \right)$.
The complexity to calculate $\widetilde{\mathbf{x}}_{g,m_g}^{\star}$ is of order $\mathcal{O}\big(2N_\mathrm{t}( 4K-2K_g)\big)$, and to construct $\mathbf{x}_{g,m_g}^{\star}$ is of order $\mathcal{O}\left(2N_\mathrm{t}^2 \right)$.
Therefore, with pre-designed $I_{g,m_g}$ and $P_{g,m_g}$, the total complexity to obtain $\mathbf{x}_{g,m_g}^{\star}$ is of order $\mathcal{O} \big( (4K-2K_g)^3+ N_\mathrm{out}[ (4K-2K_g)^2+ 6N_\mathrm{in}K_g ] + 2N_\mathrm{t}(4K-2Kg+N_\mathrm{t}) \big)$, where $N_\mathrm{out}$ denotes the number of iterations in the outer loop, and the overall computational complexity to calculate all the required precoders according to the proposed G-SLP strategy is of order $
\mathcal{O} \Big( \sum_{g=1}^G \Omega^{K_g}\big[(4K-2K_g)^3+ N_\mathrm{out}[(4K-2K_g)^2+ 6N_\mathrm{in}K_g]+ 2N_\mathrm{t}(4K-2Kg+N_\mathrm{t}) \big]\Big)$.
Besides, the complexity of obtaining a BLP solution for pre-designing $I_{g,m_g}$ and $P_{g,m_g}$ is of order $\mathcal{O}\{N_{\mathrm{iter}}N_\mathrm{t}^3\}$ with $N_\text{iter}$ representing the number of iterations in solving for $\mathbf{W}$.
It is noted that compared with the optimizations for the symbol-level precoders, the complexity of the BLP solution is negligible and thus ignored.
For the low-complexity design of traditional SLP scheme in \cite{A Li CF}, the complexity for designing each individual precoder is of order $\mathcal{O}\big(\sum_{n=0}^{n_\mathrm{max}}( 2n^2+3n+1)\big)$, where $n_\mathrm{max}$ denotes the maximum number of iterations and it increases with $K$.
Thus, the overall computational complexity for obtaining all the required precoders is of order $
\mathcal{O}\big(\Omega^{K} \sum_{n=0}^{n_\mathrm{max}}( 2n^2+3n+1)\big)$.
In addition, the sub-optimal closed-form solution in \cite{A Li CF QAM} has complexity of order $\mathcal{O}\big(16K^3 \big)$ for each precoder and $\mathcal{O}\big(16\Omega^{K}K^3 \big)$ for all possible precoders.
It can be observed that the complexity for obtaining each precoder of all above algorithms are polynomial with the number of users.
However, the number of precoders required to be designed of our proposed G-SLP strategy is remarkably less than that of the traditional SLP schemes, i.e., $N_{\mathrm{G-SLP}}=\sum_{g=1}^G \Omega^{K_g} \ll N_\mathrm{SLP}= \Omega^{K}$.
Therefore, the proposed G-SLP design algorithm is theoretically much more efficient than its counterparts.
Moreover, the numerical results of evaluating the execution time will be provided in Section \ref{sec: Simulation} to show the superiority of our proposed algorithm in complexity reduction.

\section{Low-complexity Algorithm for\\ Power Minimization Problem}
\label{sec: algorithm development for pm}

To conduct a comprehensive study on low-complexity SLP designs for MU-MISO systems, in this section we focus on the G-SLP design for the typical power minimization problem.
Specifically, with the proposed G-SLP strategy, we first formulate the design problem that minimizes the transmit power under the requirements of communication QoS and inter-group MUI suppression.
Then, we transform the optimization problem into a more tractable form following a similar procedure in the previous section.
Finally, an MM-based algorithm is developed to iteratively solve the resulting problems with closed-form solutions.
\vspace{-2ex}

\subsection{Problem Formulation for Power Minimization}

According to the G-SLP strategy proposed in Sec. III, the power minimization problem for designing $\mathbf{x}_{g,m_g}$ can be formulated as \vspace{-1ex}
\begin{subequations}
\label{GSLP PM}
\begin{align}
\min\limits_{ \mathbf{x}_{g,m_g}} & \quad \|\mathbf{x}_{g,m_g}\|_2^2\\
\label{MUI exploitation pm}
\begin{split}\text { s.t. }&~\big[ \Re\{\mathbf{h}_{g,k_g}^{H} \mathbf{x}_{g,m_g} e^{-j\angle{s_{g,m_g,k_g}}}\}-t_{g}\big] \tan \theta\\
&\quad -\big| \Im\{\mathbf{h}_{g,k_g}^{H} \mathbf{x}_{g,m_g} e^{-j\angle{s_{g,m_g,k_g}}}\}\big| \geq 0,  \quad\forall k_g,\end{split} \\
\label{interference suppress2}  &~| \mathbf{h}_{j,k_j}^{H} \mathbf{x}_{g,m_g}|^2 \leq I_{g,m_g},\quad \forall j\neq g,~\forall k_j,
\end{align}
\end{subequations}
where $t_{g}$ is the preset minimum QoS requirement for the users in the $g$-th group, $I_{g,m_g}$ is the preset maximum allowed inter-group MUI caused by $\mathbf{x}_{g,m_g}$.

Similarly, we set the inter-group MUI level $I_{g,m_g}$ as that in (\ref{Img}) based on the result of the BLP design \cite{optimal BLP}, which optimizes the block-level precoding matrix $\mathbf{W}$ to minimize the transmit power while satisfying the SINR requirement $\Gamma_0$.
The resulting optimization problem is formulated as \vspace{-1ex}
\begin{subequations}
\begin{align}
\min\limits_{\mathbf{W} }~~ & \|\mathbf{W}\|_F^2\\
\text{s.t.} ~~~&\frac{|\mathbf{h}_k^H\mathbf{w}_k|^2}{  \sum_{j=1,j\neq k}^K|\mathbf{h}_k^H\mathbf{w}_j|^2 +\sigma^2   }\geq \Gamma_0,k=1,\ldots,K,
\end{align}
\end{subequations}
which can be efficiently solved by the algorithm proposed in \cite{optimal BLP}.
In addition, for a fair comparison, the QoS requirement $t_g$ should be properly set to guarantee that the resulting SINR $\Gamma_{g,m_g}$ is no less than $\Gamma_0$, i.e., $\Gamma_{g,m_g}\geq\Gamma_0$.
In order to find the connection between $t_g$ and $\Gamma_0$, we have the following derivations
\begin{subequations}\label{eq:derive tg}
\begin{align}
\Gamma_{g,k_g}&\overset{(a)}{\geq}\frac{t_{g}^2}{ \mathbb{E}\{|\mathbf{h}_{g,k_g}^H\sum_{j=1,j\neq g}^G\mathbf{x}_{j,m_j}|^2\} + \sigma^2}\\
\label{SINR derivation}
&\overset{(b)}{=} \frac{t_{g}^2}{   \sum_{j=1,j\neq g}^G \mathbb{E}\{|\mathbf{h}_{g,k_g}^H\mathbf{x}_{j,m_j}|^2\} + \sigma^2}\\
&\overset{(c)}{\geq}\frac{t_{g}^2}{   \sum_{j=1,j\neq g}^G \mathbb{E}\{I_{j,m_j} \} + \sigma^2}\\
&\overset{(d)}{=} \frac{t_{g}^2}{   \sum_{j=1,j\neq g}^G \frac{1}{\Omega^{K_j}} \sum_{m_j=1}^{\Omega^{K_j}} I_{j,m_j} + \sigma^2}\geq\Gamma_0,
\end{align}
\end{subequations}
where (a) holds since the principle of SLP technique guarantees that the useful power (i.e., $|\mathbf{h}_{g,m_g}^H\mathbf{x}_{g,m_g}|^2$) is no less than $t_g^2$ as shown in Fig. 1, (b) holds since the precoders of each group are independent and $\mathbb{E} \{ \mathbf{x}_{g,m_g} \}=\mathbf{0},\forall g$, and (c) holds due to the interference suppression constraint (\ref{GSLP PM}c).
Therefore, to meet the SINR requirement $\Gamma_0$, the QoS requirement $t_{g}$ is set as \vspace{-1ex}
\begin{equation}\label{set t for pm}
t_{g}=\sqrt{ \Gamma_0\Bigg(  \sum_{j=1,j\neq g}^G \frac{1}{\Omega^{K_j}} \sum_{m_j=1}^{\Omega^{K_j}} I_{j,m_j} + \sigma^2 \Bigg) }.
\end{equation}

\vspace{-1ex}
\subsection{Problem Transformation for Power Minimization}

Following the procedure in Sec. \ref{problem transformation max-min}, we first approximate the interference suppression constraint (\ref{interference suppress2}) by simpler linear constraints, and relax problem (\ref{GSLP PM}) into \vspace{-1ex}
\begin{subequations}
\label{gslps pm}
\begin{align}
& \min\limits_{ \mathbf{x}_{g,m_g}}\quad \|\mathbf{x}_{g,m_g}\|_2^2 \\
\begin{split} &~\text { s.t. }~\big[ \Re\{\mathbf{h}_{g,k_g}^{H} \mathbf{x}_{g,m_g} e^{-j\angle{s_{g,m_g,k_g}}}\}- t_{g}\big] \tan \theta\\
&\quad\quad\quad\quad \pm \Im\{\mathbf{h}_{g,k_g}^{H} \mathbf{x}_{g,m_g} e^{-j\angle{s_{g,m_g,k_g}}}\} \geq 0,  \quad \forall k_g, \end{split} \\
&~\quad\quad\pm  \Re\{ \mathbf{h}_{j,k_j}^{H} \mathbf{x}_{g,m_g}\} \leq \sqrt{ I_{g,m_g} }, \quad\forall k_j,~\forall j\neq g, \\
&~\quad\quad\pm \Im\{ \mathbf{h}_{j,k_j}^{H} \mathbf{x}_{g,m_g}\} \leq \sqrt{ I_{g,m_g} },\quad  \forall k_j,~\forall j\neq g.
\end{align}
\end{subequations}
Then, using the definitions in (\ref{complex to real}) and (\ref{Hc}), we can obtain its equivalent real-valued problem as \vspace{-1ex}
\begin{subequations}
\label{pm compact}
\begin{align}
&\min\limits_{ \widetilde{\mathbf{x}}_{g,m_g}}\quad \widetilde{\mathbf{x}}_{g,m_g}^T\widetilde{\mathbf{x}}_{g,m_g}\\
&~~\text { s.t.}~~~\mathbf{A}_{g,m_g}\widetilde{\mathbf{x}}_{g,m_g}+  t_{g} \mathbf{c}_1 - \sqrt{ I_{g,m_g} } \mathbf{c}_2 \leq 0.
\end{align}
\end{subequations}
We observe that this problem is a quadratic programming with $(4K-2K_g)$ linear constraints and the Slater's conditions are satisfied.
Therefore, in order to provide a low-complexity solution, we also propose to solve its Lagrange dual problem with considering the KKT conditions.
Specifically, its Lagrangian can be expressed as
\begin{equation}
\begin{aligned}
&\mathcal{L}_3( \widetilde{\mathbf{x}}_{g,m_g},\widehat{\boldsymbol{\mu}})=\widetilde{\mathbf{x}}_{g,m_g}^T\widetilde{\mathbf{x}}_{g,m_g} \\
&\hspace{1.5 cm} + \widehat{\boldsymbol{\mu}}^T(\mathbf{A}_{g,m_g}\widetilde{\mathbf{x}}_{g,m_g}+ t_{g} \mathbf{c}_1 - \sqrt{I_{g,m_g} }   \mathbf{c}_2),
\end{aligned}
\end{equation}
where $\widehat{\boldsymbol{\mu}}\triangleq[\widehat{\mu}_1,\widehat{\mu}_2,\ldots,\widehat{\mu}_{4K-2K_g}]$ is the dual variable, $\widehat{\mu}_i \geq 0,\forall i$.
Thus, the dual problem of (\ref{pm compact}) is given by
\begin{subequations}\label{eq:pm compact dual}
\begin{align}
\label{pdual pm}
\max\limits_{\widehat{\boldsymbol{\mu}}}~&\min\limits_{ \widetilde{\mathbf{x}}_{g,m_g}}~~ \mathcal{L}_3(\widetilde{\mathbf{x}}_{g,m_g},\widehat{\boldsymbol{\mu}})\\
&\quad\text{s.t.}~\quad \widehat{\mu_i}\geq 0, \quad \forall i.
\end{align}\end{subequations}
According to the KKT conditions for optimality, the optimal solutions to problem (\ref{pm compact}) and its Lagrangian dual problem (\ref{eq:pm compact dual}) should satisfy
\begin{subequations}
\label{pm KKT}
\begin{align}
&\frac{\partial \mathcal{L}_3}{\partial \widetilde{\mathbf{x}}_{g,m_g}}= 2\widetilde{\mathbf{x}}_{g,m_g}+ \mathbf{A}_{g,m_g}^T\widehat{\boldsymbol{\mu}} =\mathbf{0}, \label{pm KKT1}\\
&  \widehat{\boldsymbol{\mu}} \odot (\mathbf{A}_{g,m_g}\widetilde{\mathbf{x}}_{g,m_g}+ t_{g}  \mathbf{c}_1 - \sqrt{ { I_{g,m_g} }}  \mathbf{c}_2) = \mathbf{0}, \label{pm KKT2}\\
&  \widehat{\mu}_i\geq0, \quad \forall i . \label{pm KKT3}
\end{align}
\end{subequations}
Based on (\ref{pm KKT1}), the optimal solution to problem (\ref{pm compact}) satisfies
\begin{equation}\label{pm x mu}
\widetilde{\mathbf{x}}_{g,m_g}^{\star}=-\frac{1}{2}{\mathbf{A}_{g,m_g}^T{\widehat{\boldsymbol{\mu}}}^{\star}},
\end{equation}
where ${\widehat{\boldsymbol{\mu}}}^{\star}$ is the optimal solution of the dual problem (\ref{eq:pm compact dual}).
Applying the result in (\ref{pm x mu}) into (\ref{eq:pm compact dual}), we can further simplify the dual problem (\ref{eq:pm compact dual}) as follows
\begin{subequations}
\begin{align}
\max\limits_{\widehat{\boldsymbol{\mu}}}~&\min\limits_{\widetilde{\mathbf{x}}_{g,m_g}}~~\mathcal{L}_3( \widetilde{\mathbf{x}}_{g,m_g},\widehat{\boldsymbol{\mu}})\\
=\max\limits_{\widehat{\boldsymbol{\mu}}}   &-\frac{1}{4}\widehat{\boldsymbol{\mu}}^T \mathbf{A}_{g,m_g} \mathbf{A}_{g,m_g}^T \widehat{\boldsymbol{\mu}}+t_{g} \widehat{\boldsymbol{\mu}}^T\mathbf{c}_1 -\sqrt{I_{g,m_g} }  \widehat{\boldsymbol{\mu}}^T\mathbf{c}_2\\
=\max\limits_{\widehat{\boldsymbol{\mu}}}   &-\frac{1}{4}\| \mathbf{V}_{g,m_g}^{\frac{1}{2}}\widehat{\boldsymbol{\mu}}\|_2^2 -\frac{1}{4}\mathbf{c}_{g,m_g}^T\widehat{\boldsymbol{\mu}},
\end{align}
\end{subequations}
where for simplicity we define $\mathbf{V}_{g,m_g}$ as that in (\ref{get V}) and
\begin{equation}
\mathbf{c}_{g,m_g}\triangleq 4\left(-t_{g}  \mathbf{c}_1 +\sqrt{I_{g,m_g} } \mathbf{c}_2\right).
\end{equation}
Therefore, taking the non-negative constraints (\ref{pm KKT3}) with respect to the variable $\widehat{ \boldsymbol{\mu}}$ into account, the dual problem (\ref{eq:pm compact dual}) is transformed into the following equivalent problem
\begin{subequations}
\label{pm simple}
\begin{align}
\min\limits_{\widehat{\boldsymbol{\mu}}}\quad & h(\widehat{\boldsymbol{\mu}})\triangleq \| \mathbf{V}_{g,m_g}^{\frac{1}{2}}\widehat{\boldsymbol{\mu}}\|_2^2 + \mathbf{c}_{g,m_g}^T\widehat{\boldsymbol{\mu}} \\
\text{s.t.} \quad & \widehat{\mu}_i\geq 0, \quad \forall i. \label{non negative constraint pm psimple}
\end{align}
\end{subequations}

It can be observed that problem (\ref{pm simple}) is a quadratic programming with $4K-2K_g$ much simpler non-negative restrictions compared with its primal problem (\ref{pm compact}).
Since this problem is very similar to problem (\ref{Psimple}) for max-min fairness design in Sec. \ref{sec:MM for max-min}, we also propose to utilize the MM method for obtaining low-complexity closed-form solutions as presented in the following subsection.
After obtaining the dual variable $\widehat{\boldsymbol{\mu}}^\star$, the solution $\widetilde{\mathbf{x}}_{g,m_g}^{\star}$ to the primal problem (\ref{pm compact}) can be calculated by (\ref{pm x mu}), and the complex-valued solution $\mathbf{x}_{g,m_g}^\star$ is constructed by (\ref{get xstar}).

\subsection{MM-based Iterative Algorithm for Power Minimization} \label{PM algorithm development}

According to the MM procedure mentioned in Sec. \ref{sec:MM for max-min}, we need to find a preferable upper-bounded surrogate function to approximate the objective function $h(\widehat{\boldsymbol{\mu}})$ in each iteration.
Based on the results in (\ref{upper bound construct}), a surrogate function of $h(\widehat{\boldsymbol{\mu}})$ is given by \vspace{-0.1cm}
\begin{equation}
h(\widehat{\boldsymbol{\mu}}|\widehat{\boldsymbol{\mu}}_t )\leq \lambda_{g,m_g}\widehat{\boldsymbol{\mu}}^T\widehat{\boldsymbol{\mu}}+\widehat{\mathbf{q}}^T\widehat{\boldsymbol{\mu}} + \widehat{\alpha},
\end{equation}
where $\widehat{\boldsymbol{\mu}}_t$ denotes the solution obtained in the $t$-th iteration, and for conciseness we define \vspace{-0.15cm}
\begin{subequations}
\label{alpha mao q mao}
\begin{align}
\widehat{\mathbf{q}}\triangleq&\  2\mathbf{V}_{g,m_g}\widehat{\boldsymbol{\mu}}_t-2\lambda_{g,m_g} \widehat{\boldsymbol{\mu}}_t+\mathbf{c}_{g,m_g},\\
\widehat{\alpha}\triangleq&\ -\widehat{\boldsymbol{\mu}}_t^T \mathbf{V}_{g,m_g} \widehat{\boldsymbol{\mu}}_t+\lambda_{g,m_g}\widehat{\boldsymbol{\mu}}_t^T\widehat{\boldsymbol{\mu}}_t.
\end{align}
\end{subequations}
Thus, the optimization problem in each iteration can be formulated as \vspace{-0.15cm}
\begin{subequations}
\label{P-minimization pm}
\begin{align}
\min\limits_{\widehat{\boldsymbol{\mu}}}\quad& \widetilde{h}( \widehat{\boldsymbol{\mu}} | \widehat{\boldsymbol{\mu}}_t )\triangleq \lambda_{g,m_g}\widehat{\boldsymbol{\mu}}^T\widehat{\boldsymbol{\mu}}+\widehat{\mathbf{q}}^T\widehat{\boldsymbol{\mu}}\\
\text{s.t.} \quad & \widehat{\mu}_i\geq 0, \quad \forall i,  \label{non negative constraint pm minimization}
\end{align}
\end{subequations}
whose optimal closed-form solution can be easily calculated as \vspace{-0.15cm}
\begin{equation}
\label{get mu hat}
\widehat{\boldsymbol{\mu}}^{\star}=\max\Big\{ \mathbf{0},-\frac{ \widehat{\mathbf{q}}} {2\lambda_{g,m_g}  } \Big\}.
\end{equation}

\begin{algorithm}[t]\begin{small}
  \caption{MM-based G-SLP Design for Power Minimization Problem (\ref{gslps pm})}
  \label{Algorithm for pm}
  \begin{algorithmic}[1]
    \REQUIRE $\mathbf{H}_g$, $\mathbf{s}_{g,m_g}$, $I_{g,m_g}$, $t_{g}$, $\epsilon_3$.
    \ENSURE  $\mathbf{x}_{g,m_g}^{\star}$.
    \STATE {Initialize $t := 0$, $\widehat{\boldsymbol{\mu}}_{t}:=\mathbf{0}$.}
    \STATE {Calculate $\mathbf{A}_{g,m_g}$ by (\ref{Hc}), $\mathbf{V}_{g,m_g}$ by (\ref{get V}), and $\lambda_{g,m_g}$;}
    \REPEAT
    \STATE {Calculate $\widehat{\mathbf{q}}$ by (\ref{alpha mao q mao}a);}
    \STATE {Update $\widehat{\boldsymbol{\mu}}_{t+1}$ by (\ref{get mu hat});}
    \STATE {$t := t + 1$;}
    \UNTIL {$\|\widehat{\boldsymbol{\mu}}_{t}-\widehat{\boldsymbol{\mu}}_{t-1}\|_2/\|\widehat{\boldsymbol{\mu}}_{t-1}\|_2 \leq \epsilon_3$.}
    \STATE{$\widehat{\bm{\mu}}^\star = \widehat{\bm{\mu}}_t$.}
    \STATE {Calculate $\widetilde{\mathbf{x}}_{g,m_g}^{\star}$ by (\ref{pm x mu});}
    \STATE {Construct $\mathbf{x}_{g,m_g}^{\star}$ by (\ref{get xstar}).}
  \end{algorithmic}\end{small}
\end{algorithm} 

\vspace{-1ex}
\subsection{Algorithm Summary and Complexity Analysis}

With above derivations, the proposed G-SLP design algorithm for power minimization problem is straightforward and summarized in Algorithm \ref{Algorithm for pm}, where $\epsilon_3$ is a parameter to judge the convergence.
In summary, we iteratively update the dual variable $\widehat{\bm{\mu}}$ with a closed-form solution (\ref{get mu hat}) until convergence, then recover the solution to the primal problem (\ref{pm compact}) by (\ref{pm x mu}) and finally construct the complex-valued solution by (\ref{get xstar}).

The computational complexity of steps 1-2 is of order $\mathcal{O}( (4K-2K_g)^3 )$.
In each iteration, updating $\widehat{\mathbf{q}}$ and $\widehat{\boldsymbol{\mu}}_t$ has complexity of order $\mathcal{O}( (4K-2K_g )^2)$.
The complexity to calculate $\widetilde{\mathbf{x}}_{g,m_g}^{\star}$ is of order $\mathcal{O}(   2N_\mathrm{t}(4K-2K_g ))$, and to construct $\mathbf{x}_{g,m_g}^{\star}$ is of order $\mathcal{O}(2N_\mathrm{t}^2)$.
Therefore, the total complexity of Algorithm \ref{Algorithm for pm} is of order $\mathcal{O}\big(  (4K-2K_g)^3+2N_\mathrm{t}(4K-2K_g +N_\mathrm{t})+N_\mathrm{tot} (4K-2K_g )^2  \big)$, where $N_\mathrm{tot}$ denotes the number of iterations.
The overall complexity of designing all the required precoders is of order $
\mathcal{O}\big(\sum_{g=1}^G \Omega^{K_g} [ (4K-2K_g)^3+2N_\mathrm{t}(4K-2K_g +N_\mathrm{t})+
N_\mathrm{tot} (4K-2K_g)^2 ] \big)$.
In addition, the overall complexity of the sub-optimal closed-form solution of the traditional SLP scheme for power minimization problem in \cite{PM ICF} is of order $\mathcal{O}(\Omega^K N_\mathrm{t}(K+L_1^2+L_2^2))$, where $L_1$ and $L_2$ are design parameters and $L_2\leq L_1 \ll2K$.
Similarly, we can observe that the proposed G-SLP design algorithm is theoretically much more efficient than the
traditional SLP scheme for the power minimization problem.

\section{Simulation Results} \label{sec: Simulation}

In this section, we provide numerical experiments to illustrate the advantages of the proposed G-SLP strategy and the effectiveness of the proposed design algorithms for both max-min fairness and power minimization problems.
We assume that the BS has a uniform linear array (ULA) with half-wavelength antenna spacings, the transmitted symbols are independently selected from QPSK constellation, i.e., $\Omega=4$, and the $K$ users are uniformly divided into $G$ groups, $G = 2, 3$.
The noise power of users is set as $\sigma^2=10$dBm. The parameters $\epsilon_1$, $\epsilon_2$, and $\epsilon_3$ are all set as $1e-6$.
All the simulations are carried out using Matlab R2020b on a PC with an Intel Core i7-10700F CPU and 32GB of RAM.

\subsection{Max-min Fairness Case}

\begin{figure}[t]
\centering
\includegraphics[width=3.5 in]{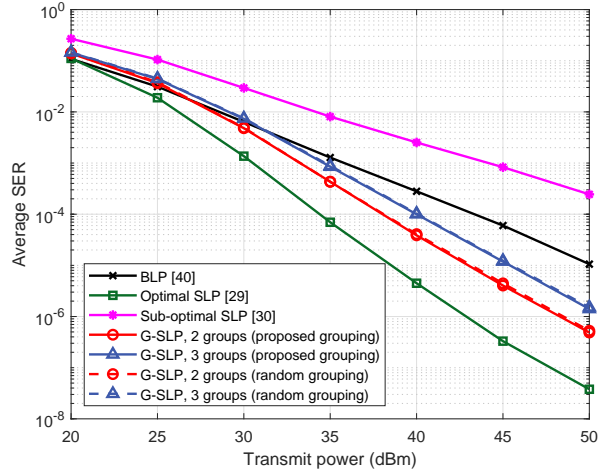}
\caption{Average SER versus transmit power $P$ under uncorrelated channel ($N_\mathrm{t}=K=12$).} \label{SB1}
\end{figure}

In this subsection, we evaluate the performance of the proposed G-SLP design algorithm in Sec. \ref{sec: algorithm development for max-min} for the max-min fairness problem.
The traditional SLP solutions obtained by the optimal iterative algorithm \cite{A Li CF} (denoted as ``Optimal SLP \cite{A Li CF}'') and the sub-optimal closed-form algorithm \cite{A Li CF QAM} (denoted as ``Sub-optimal SLP \cite{A Li CF QAM}''), and the BLP solution \cite{optimal BLP} (denoted as ``BLP \cite{optimal BLP}''), are also included for comparison purposes.
For fairness, the transmit power at each symbol time-slot of all schemes is scaled to $P$, i.e., $P_m=P_0=P$, $\forall m$.

We first illustrate the SER performance as a function of the transmit power $P$ in Fig. 3, where the schemes with random grouping are also included for comparison.
We assume that the system is fully loaded with $N_\text{t} = K = 12$ and adopts the uncorrelated Rayleigh fading channel model, with which the elements of the channel vectors have independent and identically complex Gaussian distribution.
It can be noticed that both the optimal SLP and proposed G-SLP schemes achieve better performance than the BLP scheme when the transmit power is relatively high owing to the MUI exploitation of SLP technique.
Moreover, this advantage appears more obviously as the transmit power increases since the MUI becomes a dominant factor that affects the SER performance and larger transmit power can provide more MUI to be exploited.
In addition, while the proposed G-SLP schemes achieves much better performance than its counterpart of the sub-optimal SLP design proposed in \cite{A Li CF QAM}, we observe that the SER performance of the G-SLP schemes are worse than the traditional SLP schemes.
This phenomenon results from the design principle of G-SLP, which only utilizes the intra-group MUI while suppresses the inter-group MUI. The performance of the 3-group scheme is worse than that of the 2-group scheme since more groups can further reduce the complexity with smaller number of user in each group, but have less intra-group MUI to be exploited.
However, considering the superiority of G-SLP schemes in complexity reduction as shown in Fig. \ref{SBMU2}, the performance losses are acceptable especially when implementing in a dense-user system.
We also see that compared to the schemes with random grouping, the proposed G-SLP schemes only offer very slight performance improvement.
The reason for this phenomenon lies in the considered uncorrelated channel model, with which the interference between any two users is generally at a similar level and thus the impact of the grouping strategy is marginal.

\begin{figure}[t]
\centering
\includegraphics[width=3.5 in]{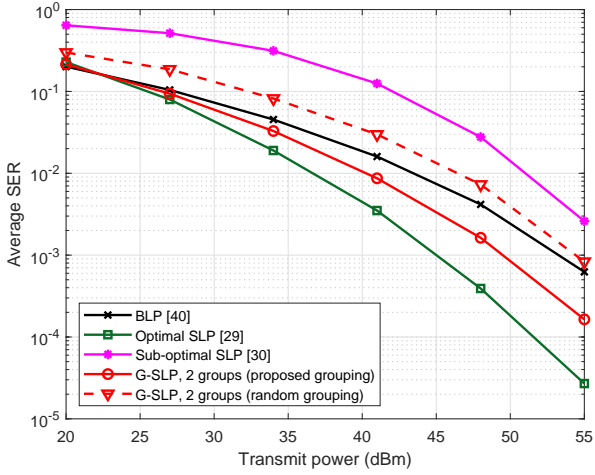}
\caption{Average SER versus transmit power $P$ under correlated channel ($N_\mathrm{t}=32$, $K=12$).} \label{SB2}
\end{figure}

Next, in order to illustrate the importance of grouping strategy, we consider the correlated channel model, which generates the channel vectors using the correlation matrix.
The well-known one-ring model \cite{one-ring model}, \cite{one-ring model 2} is adopted  for generating correlated channels, which assumes that a ring of scatterers around the users and there is no scattering close to the BS.
In particular, assuming that a ULA at the BS formed by $N_\mathrm{t}$ directional radiating elements is placed at the origin along the $y$-axis, each user is located at azimuth angle $\theta$ and the departure signals have a small angular spread $\Delta$ around it from the BS perspective.
Specifically, the channel vector $\mathbf{h}_k$ is generated by
\begin{subequations}\begin{align}
\mathbf{h}_k & = {\mathbf{R}_k^{\frac{1}{2}} }\widehat{\mathbf{h}}_k,\\
\mathbf{R}_k(m,p)&=\frac{1}{2\Delta}\int_{-\Delta+\theta}^{\Delta+\theta} e^{-j\pi (m-p)sin(\beta) }d\beta,
\end{align}\end{subequations}
where $\widehat{\mathbf{h}}_{k}\backsim\mathcal{CN}(\mathbf{0},\mathbf{I}_{N_\mathrm{t}})$ and $\mathbf{R}_k $ is the correlation matrix.
Fig. \ref{SB2} illustrates the SER performance under the considered the correlated channel model.
The BS is equipped with $N_\mathrm{t}=32 > K$ antennas to serve $K=12$ users in order to combat the high correlations.
To show the effect of channel correlations on the system performance, we assume a highly correlated environment where the angular spread is set as  $\Delta=8^{\circ}$, and the azimuth angle $\theta$ of one half of the users is randomly generated in $[-60^{\circ}-\theta_\delta, -60^{\circ}+\theta_\delta]$ and the other half of the users in $[60^{\circ}-\theta_\delta, 60^{\circ}+\theta_\delta]$ with $\theta_\delta = 5^\circ$.
Therefore, all users can be equally divided into 2 groups for implementing the proposed G-SLP scheme.
In addition to the same performance relationship as shown in Fig. \ref{SB1}, a notable performance improvement introduced by the proposed grouping strategy can be observed from Fig. \ref{SB2}.
This is because the interference between different users varies significantly under the correlated channel model, which makes the grouping strategy very crucial in elaborating MUI.
These results demonstrate the effectiveness of the proposed scheme for both uncorrelated and correlated channels, and especially the advantage of the proposed grouping strategy for correlated channels.

\begin{figure}[t]
\centering
\includegraphics[width=3.5 in]{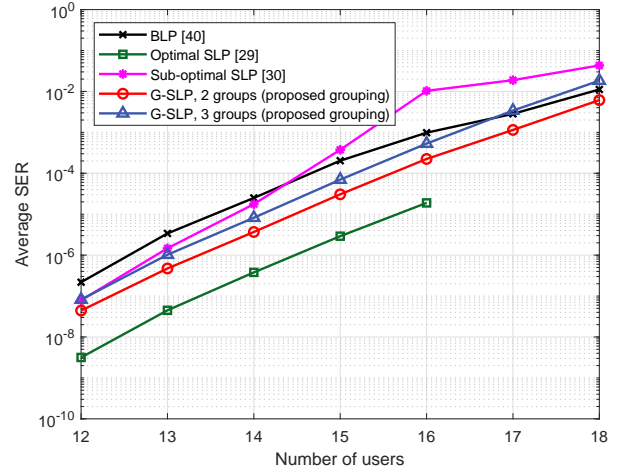}
\caption{Average SER versus the number of users $K$ under uncorrelated channel ($N_{\text{t}} = 16$, $P=35\text{dBm}$). } \label{SBMU1}
\vspace{-0.2 cm}
\end{figure}

\begin{figure}[t]
\centering
\includegraphics[width=3.5 in]{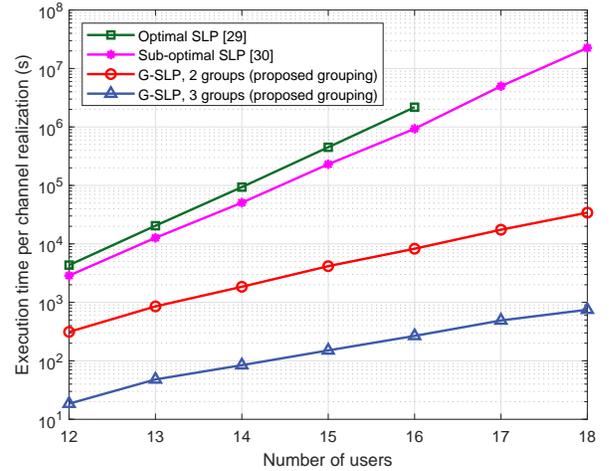}
\caption{Total execution time versus the number of users $K$ under uncorrelated channel ($N_{\text{t}} = 16$, $P=35\text{dBm}$).}\label{SBMU2}
\end{figure}

Then, we present the SER performance as a function of the number of users in Fig. \ref{SBMU1} with uncorrelated channel, where the number of transmit antennas is set as $N_\text{t}=16$ and the transmit power is set as $P=35$dBm.
Since the optimal closed-form solution \cite{A Li CF} was designed only for the scenarios that $K\leq N_\text{t}$, there are no simulation results of the ``Optimal SLP'' scheme when $K>16$.
We observe that while the performance of the sub-optimal closed-form solution of SLP in \cite{A Li CF QAM} is greatly affected by the number of users, the
proposed G-SLP schemes can maintain satisfactory performance even when the number of users is larger than that of transmit antennas.
These results demonstrate the robustness of our proposed approach.
Next, the corresponding total execution time for designing all the required precoders during a channel coherent time versus the number of users is shown in Fig. \ref{SBMU2}.
While the theoretical analyses in Sec. IV-C demonstrate the effectiveness of our proposed scheme, these numerical results of the practical execution time validate the notable advantages over traditional SLP schemes from another more straightforward perspective.
It is also noted that the relationship between the execution times required by different algorithms is consistent with the analytical results.
Furthermore, as the number of users increase, the advantage of our proposed G-SLP algorithm in reducing complexity is more remarkable while the corresponding performance gap shown in Fig. \ref{SBMU1} maintains at a similar level.
In addition, we observe that the G-SLP scheme with less groups has lower complexity at the price of certain performance loss, which reveals the flexibility of the proposed G-SLP strategy in balancing complexity and performance.

\begin{figure}[t]
\centering
\includegraphics[width=3.5 in]{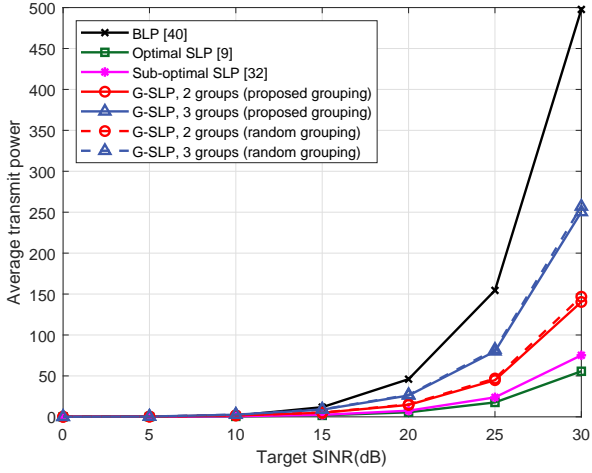}
\caption{Transmit power versus SINR under uncorrelated channel ($N_\mathrm{t}=K=12$). } \label{PM1}
\end{figure}

\begin{table}[t]
\small
\caption{\label{SER for PM1}Average SER versus SINR under uncorrelated channel ($N_{\text{t}}=K=12$).}
\centering
\begin{tabular}{p{1.9cm} p{1cm} p{1cm}  p{1cm} p{1cm} p{1cm}}
\toprule
Target SINR & 5 & 7  & 9 & 11 \\
\hline
$\mathrm{SER}_
{\text{BLP \cite{optimal BLP}} }$   & $7.4e-2$ & $2.4e-2$  & $4.3e-3$& $3.1e-4$   \\
$\mathrm{SER}_{\text{SLP \cite{SLP performance}}}$   & $5.5e-2$ & $1.8e-2$  & $3.5e-3$& $2.8e-4$  \\
$\mathrm{SER}_{\text{SLP \cite{PM ICF}}}$  & $6.1e-2$ & $2.0e-2$  & $3.9e-3$& $3.1e-4$   \\
$\mathrm{SER}_{\text{G-SLP, 2 groups}}^{\text{proposed grouping}}$  & $5.0e-2$ & $1.3e-2$  & $1.7e-3$ & $1.1e-4$   \\
$\mathrm{SER}_{\text{G-SLP, 3 groups}}^{\text{proposed grouping}}$ & $5.5e-2$ & $1.5e-2$  & $1.9e-3$ & $1.1e-4$   \\
$\mathrm{SER}_{\text{G-SLP, 2 groups}}^{\text{random grouping}}$  & $5.0e-2$ & $1.2e-2$  & $1.7e-3$ & $1.0e-4$   \\
$\mathrm{SER}_{\text{G-SLP, 3 groups}}^{\text{random grouping}}$ & $5.6e-2$ & $1.5e-2$  & $1.9e-3$ & $1.1e-4$   \\
\bottomrule
\end{tabular}
\end{table}

\subsection{Power Minimization Case}

In this subsection, we illustrate the performance of the proposed G-SLP design algorithm in Sec. \ref{sec: algorithm development for pm} for the power minimization problem.
For comparison purpose, we also include the optimal iterative SLP solution in \cite{SLP performance}, the state-of-the-art sub-optimal closed-form SLP solution in \cite{PM ICF}, and the BLP solution in \cite{optimal BLP}, which are denoted as ``Optimal SLP \cite{SLP performance}'', ``Sub-optimal SLP \cite{PM ICF}'', and ``BLP \cite{optimal BLP}'', respectively.

\begin{figure}[t]
\centering
\includegraphics[width=3.5 in]{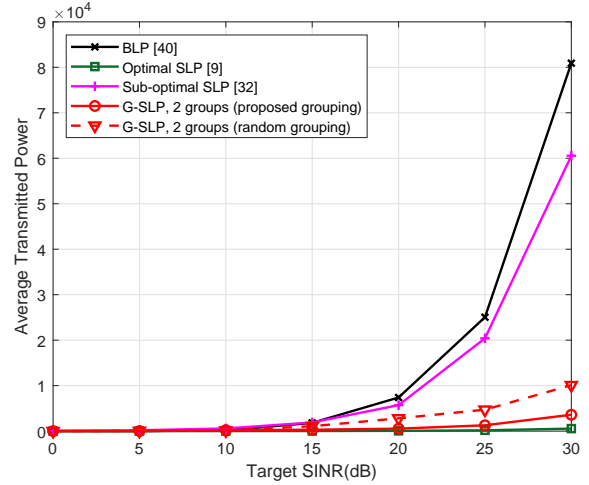}
\caption{Transmit power versus SINR under correlated channel ($N_\mathrm{t}=32$, $K=12$). } \label{PM2}
\end{figure}

We first plot the average transmit power versus the SINR requirement under uncorrelated channel in Fig. \ref{PM1}, where the channel settings are the same as that in Fig. 3.
Similar conclusions as that of the max-min fairness case can be drawn from Fig. 3: Our proposed G-SLP schemes have significant improvements over traditional BLP method especially when the transmit power or SINR is large; there exist acceptable performance losses compared with traditional SLP scheme; and the more groups will cause the greater performance loss.
The corresponding SERs of these schemes in Fig. \ref{PM1} are also provided in Table \ref{SER for PM1} to verify the fairness of our settings.
From Table \ref{SER for PM1} we observe that the SERs of the proposed G-SLP schemes are lower than that of the traditional SLP schemes and the BLP schemes, which verifies that the QoS constraints of our schemes are stricter than the others as derived in (\ref{eq:derive tg}).
Therefore, if all the schemes provide the same SER performance, the power gap between the proposed G-SLP schemes and the traditional SLP scheme will be smaller, while the power gap between the proposed G-SLP schemes and the BLP scheme will become larger compared with that in Fig. \ref{PM1}.
This result further illustrates that the proposed G-SLP schemes and the developed design algorithm have very satisfactory power minimization performance.

Then, in Fig. \ref{PM2} we repeat the same simulation by considering the correlated channel as that in Fig. \ref{SB2}.
Similarly, significantly better performance offered by the proposed grouping strategy can be observed.
Furthermore, we notice that our proposed G-SLP strategy exhibits acceptable performance losses under the uncorrelated channels but has much better performance under the correlated channels compared with its counterpart of the closed-form suboptimal solution in \cite{PM ICF}.

\begin{figure}[t]
\centering
\includegraphics[width=3.5 in]{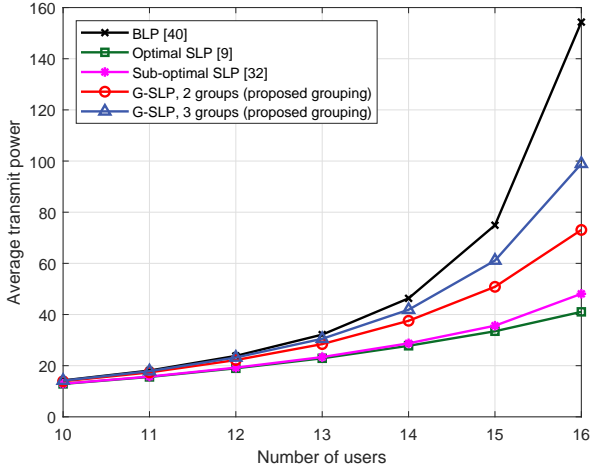}
\caption{Transmit power versus the number of users $K$ under uncorrelated channel ($N_{\text{t}} = 17$, $\text{target\ SINR}=30\text{dB}$). }\label{MUPM1}
\vspace{-0.2 cm}
\end{figure}

\begin{figure}[t]
\centering
\includegraphics[width=3.5 in]{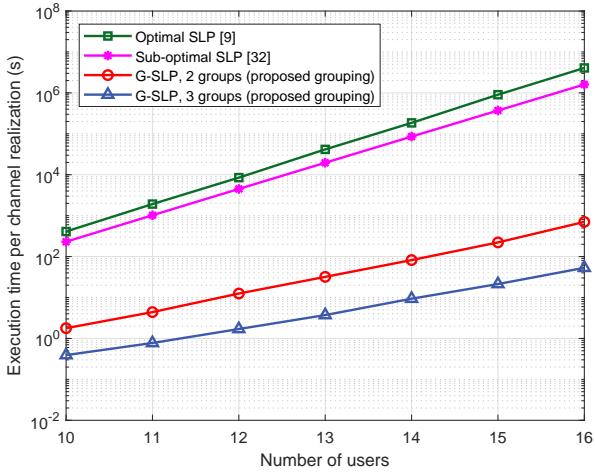}
\caption{Total execution time versus the number of users $K$ under uncorrelated channel ($N_{\text{t}} = 17$, $\text{target\ SINR}=30\text{dB}$). }\label{MUPM2}
\vspace{-0.2 cm}
\end{figure}

Finally, the average transmit power and the corresponding total execution time as a function of the number of users for different precoding methods are respectively presented in Figs. \ref{MUPM1} and \ref{MUPM2}, where the number of transmit antennas is set as $N_\mathrm{t} = 17$ and the SINR requirement is set as 30dB.
We can also observe that our proposed G-SLP schemes maintain very satisfactory power minimization performance with dramatically less execution time.
This advantage becomes more obvious when the user number $K$ increases, for example, when $K=16$ the execution time of G-SLP is reduced by about $10^4$ times for 2-group G-SLP and $10^5$ times for 3-group G-SLP. Therefore, our proposed G-SLP scheme is more appealing than traditional SLP in practical dense-user systems since it can provide satisfactory performance with affordable design complexity.

\section{Conclusions}

In this paper, we introduced a novel G-SLP strategy to dramatically reduce the computational complexity of SLP designs.
With the proposed G-SLP strategy, efficient algorithms for max-min fairness and power minimization problems were respectively developed by leveraging the Lagrangian dual, KKT conditions, and the MM method.
Extensive simulation results verified that the proposed algorithms have remarkable superiority in complexity reduction with acceptable performance loss compared with the traditional SLP schemes.
Meanwhile, considerable performance improvements over BLP schemes are offered by the proposed efficient G-SLP design algorithms.
For practical dense-user system, our proposed G-SLP scheme is more appealing than traditional SLP owing to its excellent balance between system performance and design complexity.


\end{document}